\documentclass[a4paper,final]{article}
\usepackage{latexsym,amsmath,amssymb,graphicx,showkeys}

\newcommand{\scri}{\mathcal{J}}
\newcommand{\RR}{\mathbf{R}}
\newcommand{\CC}{\mathbf{C}}
\newcommand{\cA}{\mathcal{A}}
\newcommand{\cB}{\mathcal{B}}
\newcommand{\cN}{\mathcal{N}}
\newcommand{\cS}{\mathcal{S}}
\newcommand{\D}{\mathcal{D}}

\newcommand{\del}{\partial}
\newcommand{\2}{\frac12}

\numberwithin{equation}{section}

\title {Numerical treatment of the hyperboloidal initial value problem
for the vacuum Einstein equations\\
II. The evolution equations}
\author{J\"org Frauendiener}

\begin{document}
\maketitle

\begin{abstract}
  This is the second in a series of articles on the numerical solution
  of Friedrich's conformal field equations for Einstein's theory of
  gravity. We will discuss in this paper the numerical methods used to
  solve the system of evolution equations obtained from the conformal
  field equations. In particular we discuss in detail the choice of
  gauge source functions and the treatment of the boundaries. Of
  particular importance is the process of ``radiation extraction''
  which can be performed in a straightforward way in the present
  formalism.
\end{abstract}

\section{Introduction}
\label{sec:intro}

In article \cite{jf-1997-2}  we have
presented the conformal field equations explicitly in a form suitable
for solving them numerically. In a brief summary, we have derived the
conformal field equations in the space spinor formalism which is well
suited to perform the $3+1$ split because the evolution equations come
out in a symmetric hyperbolic form (under appropriate assumptions on
the free gauge source functions) almost automatically. We have also
discussed in \cite{jf-1997-2} the further assumption of a hypersurface
orthogonal symmetry which has been made to simplify the
implementation.  Fixed points of a continuous symmetry usually lead to
coordinate singularities which have to be treated specially in any
finite difference method. Therefore, we followed a suggestion of
B.~G.~Schmidt \cite{Schmidt-1996} to require that there be no fixed
points which has the unphysical consequence that the global topology
of space-time is $T^2\times \RR^2$. However, since the emphasis of
this project lies in studying the effectiveness of radiation
extraction from the numerically generated space-times and since these
are local methods this is not a serious disadvantage.

In section \ref{sec:num} we first present the numerical method, a
Lax-Wendroff method in two dimensions and the procedure for choosing
the time-step dynamically in order to enforce the CFL condition for
stability of the algorithm. In section \ref{sec:bndry} we show how the
boundary can be treated. This is an essential part of any numerical
scheme because if the boundary conditions are non-physical one has to
live with the fact that the numerical solution probably differs quite
significantly in the domain of influence of the boundary from what one
expects it to be there. In the present approach this problem can be
avoided because the boundary is entirely outside the physical
space-time. From the causal properties of the evolution equations it
is at least plausible that null infinity $\scri$ which is a
characteristic surface for the differential equations also numerically
acts as a barrier for perturbations generated in the unphysical
space-time.

The section \ref{sec:gauge} is devoted to a discussion of the various
gauge source functions. This is an important subject also in the
conventional treatments of the Einstein equations because it is not
clear what the implications are if one chooses e.g. the lapse function
and the shift vector in that way and not in another. The existence of
$\scri$ imposes some questions concerning the resolution of features
inside the physical space-time during the course of the
computation. But these questions can be solved completely
satisfactorily by making use of a certain choice of shift vector which
is forced by the structure of conformal field equations.

Finally, in section \ref{sec:massrad} we discuss the procedure of
``radiation extraction''. By this term we mean the determination of
certain asymptotic quantities which in a well defined sense
characterise the gravitational radiation generated inside the physical
space-time and escaping out to infinity. Here, this is a well defined
procedure which involves finding the zero-set of the  conformal
factor, the interpolation of the field variables and a frame
transformation to a well defined frame which is adapted to the
geometry of $\scri$. Another issue discussed in this section is the
determination of the Bondi mass. Here, due to the different global
topology, the situation is different from the physical one in that one
cannot find a Bondi four-momentum but only a Bondi scalar.

We tested the code by using exact solutions to first provide initial
data for all the variables and then to compare the computed variables
with the analytic ones. We also computed the radiative quantities from
the exact solutions for comparison with the numerically determined
ones to check the radiation extraction method. 

As in \cite{jf-1997-2} we use the conventions of
\cite{PenroseRindlerII} throughout.

\section{The numerical evolution scheme}
\label{sec:num}
The evolution part of the conformal field equations presented in
\cite[section \ref{sec:eqns}]{jf-1997-2} with the symmetry reduction
described in \cite[section \ref{sec:symm}]{jf-1997-2} are to be solved
numerically. To this end we set up a two-dimensional grid with
coordinates $u,v$. It is assumed that the fields are periodic in $u$
with period 2 so that we need to impose periodic boundary conditions
at the surfaces $u=\pm 1$. The coordinate $v$ takes values in the
interval $\left[-v_0,v_0\right]$. The question of boundary conditions
at the surfaces $v=\pm v_0$ is rather delicate from the numerical
point of view and will be discussed in detail in section
\ref{sec:bndry}.

Having set up the grid we need to obtain solutions of the constraint
equations in order to initialise the fields. This should be done by
solving the constraint equations numerically given the appropriate
free data and boundary data. However, since we are here concerned
mainly with the evolution algorithm we will simply take these initial
values from appropriately adapted exact solutions which have recently
been pointed out by Schmidt \cite{Schmidt-1996}. In particular, we
take the rescaled A3 solution and one other solution from this class
(see appendix \ref{sec:exsol}) in various gauges as our test case.

As our numerical method for solving the evolution equations we choose
finite difference schemes which are second order accurate in both time
and space. In particular, we use the leapfrog and the Lax-Wendroff
schemes both extended in a straightforward manner to two space
dimensions. Of course, various other methods could have been employed.
It soon becomes apparent that the leapfrog method is not a viable
choice in our case. Since the conformal field equations form a
quasi-linear symmetric hyperbolic system it follows that the
characteristics which determine the evolution of the fields depend on
the solution itself. Or, to put it differently, the wave parts of the
fields propagate along the light cone which is determined by the
metric which, in turn, is evolved by the field equations. Since the
methods we employ are explicit we need to make sure that they remain
stable by controlling the size of the time step $\Delta t$ during the
evolution. However, changing the time step dynamically cannot be done
with the leapfrog method without loosing the second order accuracy in
time. Hence, we will focus here exclusively on the Lax-Wendroff
method.

The equations in the general form given in \cite[section
\ref{sec:eqns}]{jf-1997-2} are manipulated using the \texttt{NPspinor}
package \cite{McLenaghan-1987} of \texttt{Maple}, extended to include
the space spinor formalism. The equations are expanded into components
using the decomposition into irreducibles for each spinor field. We
should point out that the equations when written in components turn
into, in general, complex equations for complex variables. Due to the
reality properties of the spinor fields these equations come either in
complex conjugate pairs or as real equations. This fact reduces the
number and the complexity of the equations.  The symmetry conditions
given in \cite[section \ref{sec:symm}]{jf-1997-2} are used to simplify
them. \texttt{Maple} is also used to test the equations thus obtained
quite extensively in various ways:
\begin{itemize}
\item They are checked against hand calculations in simple enough
  cases.
\item Inserting exact solutions into the equations should result in
  identities. These solutions are obtained also with the help of
  \texttt{Maple} using different routines by conformally transforming
  simple vacuum solutions of Einstein's equations with arbitrary
  conformal factors.
\item The most important test, however, is the fact that the evolution
  equations have to propagate the constraints. This property was
  verified for the full expanded evolution equations and constraints
  using \texttt{Maple}.
\end{itemize}

The two-dimensional Lax-Wendroff scheme is a straightforward
generalisation of the one dimensional case. For a quasi-linear
equation of the form
\begin{equation}
  f_t = A(t,u,v,f) f_u + B(t,u,v,f) f_v + E(t,u,v,f)
\end{equation}
it can be characterised as follows. Define the following operators
acting on a grid function $f_{i,j}$
\begin{align}
  \mu_1[f]_{i,j}&=\2 \left( f_{i-\2,j} + f_{i+\2,j} \right),&
  \mu_2[f]_{i,j}&=\2 \left( f_{i,j-\2} + f_{i,j+\2} \right), \\
  D_1[f]_{i,j}&=\left( f_{i+\2,j} - f_{i-\2,j} \right),&
  D_2[f]_{i,j}&=\left( f_{i,j+\2} - f_{i,j-\2} \right).
\end{align}
Then the 2D-Lax-Wendroff scheme consists of the four steps
\begin{align*}
  \bar{f}^n_{i+\2,j+\2} &\leftarrow 
\2 \mu_1 \left[\mu_2 \left[f^n \right]\right]_{i+\2,j+\2},\\ 
   f^{n+\2}_{i+\2,j+\2} &\leftarrow
   \bar{f}^n_{i+\2,j+\2} 
   + \2 C_u A^n_{i+\2,j+\2}
       \mu_2\left[D_1\left[f^n\right]\right]_{i+\2,j+\2}\\
   &+ \2 C_v B^n_{i+\2,j+\2}
       \mu_1\left[D_2\left[f^n\right]\right]_{i+\2,j+\2}
   + \2 \Delta t\, E^n_{i+\2,j+\2},\\ 
   \bar{f}^{n+\2}_{i,j} &\leftarrow 
   \2 \mu_1\left[\mu_2 \left[f^{n+\2} \right]\right]_{i,j},\\ 
   f^{n+1}_{i,j} &\leftarrow 
   f^n_{i,j} 
   + C_u A^{n+\2}_{i,j} 
       \mu_2\left[D_1\left[f^{n+\2}\right]\right]_{i,j}\\
   &+ C_v B^{n+\2}_{i,j} 
       \mu_1\left[D_2\left[f^{n+\2}\right]\right]_{i,j}
   + \Delta t\, E^{n+\2}_{i,j}.
\end{align*}
where $C_u=\frac{\Delta t}{\Delta u}$ and $C_v=\frac{\Delta t}{\Delta
v}$ and where $A^n_{i+\2,j+\2} = A(t_n,u_i+\2 \Delta u,v_j+\2 \Delta
v,\bar{f}^n_{i+\2,j+\2})$ and $A^{n+\2}_{i,j}=
A(t_n,u_i,v_j,\bar{f}^{n+\2}_{i,j})$. The further generalisation of
this scheme to three dimensions is straightforward. However, it
becomes rather inefficient and it is here where probably operator
splitting methods should be used. The complete discretisation of the
equations was also carried out symbolically.

\begin{figure}[htbp]
  \begin{center}
    \leavevmode
    \includegraphics[height=5cm]{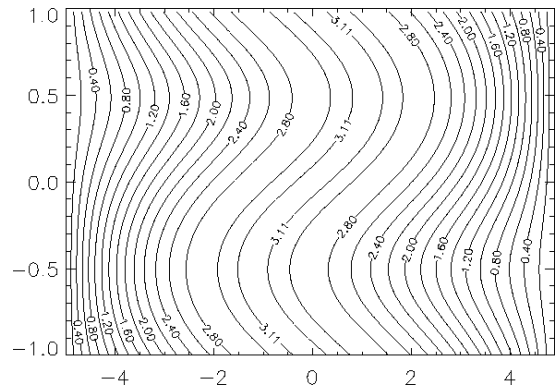}
    \caption{The orbits of the additional Killing vector $\del_u$}
    \label{fig:coord}
  \end{center}
\end{figure}

\begin{figure}[htbp]
  \begin{center}
    \leavevmode
    \includegraphics[height=5cm]{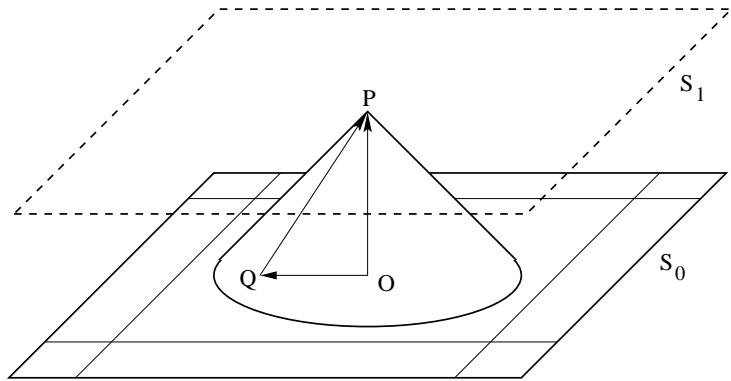}
    \caption{The local geometry for the timestep criterion}
    \label{fig:timestep}
  \end{center}
\end{figure}

The exact solutions described in appendix~\ref{sec:exsol} have been
used to provide the initial data and (in some cases) those boundary
data which can be specified freely. It is clear from the form of the
metric that these solutions have two space-like Killing vectors. The
Killing vector $\del_y$ is taken to be the one that is factored out by
the symmetry reduction (see \cite[section
\ref{sec:symm}]{jf-1997-2}). The metric functions are independent of
$x$ and $y$. If we choose $z$ and $x$ to correspond to the two
remaining coordinates $v$ and $u$ respectively then the code is
essentially a one-dimensional code. In order to test the
two-dimensional performance of the code we therefore have to ``warp''
the coordinates. Thus, we put
\begin{equation}
  \label{eq:coordtrans}
  x = u, \qquad z = v-\alpha (v_0^2-v^2)\sin(\pi u).
\end{equation}
We choose $\alpha$ in such a way that this transformation is bijective
in the range $u\in [-1,1], v\in [-v_0,v_0]$. In this coordinate system
the orbits of the second Killing vector are distorted and not aligned
with the grid, see Fig.~\ref{fig:coord}. All computations which are
presented here have been performed with these warped coordinates.
We will now describe the criterion to determine the maximal time step
$\Delta t$ possible to evolve from an initial time level $\cS_0$ at
time $t=t_n$ to the next time level $\cS_1$ at time $t=t_n+ \Delta
t$.  This is not specific to the Lax-Wendroff method but can be used
with any explicit evolution scheme. The Courant-Friedrichs-Lewy
condition \cite{CourantFriedrichsLewy-1928} can be phrased as stating
that ``the numerical domain of dependence should enclose the
analytical domain of dependence''. Now consider a point $P$ in the
future time level (cf. fig. \ref{fig:timestep}).
Its numerical domain of dependence consists of the points at time
$t_n$ which are used to compute the field values at $P$. They lie
within a rectangular area $\mathcal{R}$ bounded by coordinate lines
$u=u_\pm$ and $v=v_\pm$. The analytic domain of dependence is given by
the intersection of the backward light cone of $P$ with the time slice
$\cS_0$. The maximal allowed time-step $\Delta t$ is therefore at most
so big that the light cone just touches the boundary of
$\mathcal{R}$. To obtain a formula for the maximal $\Delta t$ we
analyse this situation to first order or, what is the same thing, in
Minkowski space. Then the time levels are planes and the light cone is
a true cone, the null cone of the point $P$. Let $O$ be the point in
$\cS_0$ which is ``straight below'' $P$ in the sense that
$\overrightarrow{OP}$ is a multiple of the normal vector $t^a$ of
$\cS_0$. We take $O$ as the origin. Let $Q$ be the point in $\cS_0$
with the same spatial coordinates as $P$ so that $\overrightarrow{QP}
= \Delta t \del_t$. The equation for the plane $\cS_1$ is given by
$\langle dt,x\rangle = \langle dt, \overrightarrow{QP} \rangle =
\Delta t$ so that $\overrightarrow{OP}^a=N t^a\Delta t$. Then
$\overrightarrow{OQ}$ is proportional to the shift vector
$\overrightarrow{OQ} = - N \Delta t T^i \del_i$. The equation for the
null cone of $P$ is
\begin{equation}
  g_{ab}\left(x^a - \overrightarrow{OP}^a\right)\left(x^b -
    \overrightarrow{OP}^b\right)=0 
\end{equation}
and its intersection $E$ with $\cS_0$ is given by all points $x^a$
which obey the equations
\begin{equation}
  t_ax^a=0,\qquad x^a x_a = 2 N^2 \Delta t^2 = 0.
\end{equation}
Now we take any plane $H$ orthogonal to $\cS_0$, whose equation is
$\omega_a x^a = s$ for some $\omega$ with $\omega_a t^a=0$ and
arbitrary $s$. Suppose that $H$ touches $E$ in a point $X$. Then, at
$X$ the following equations hold:
\begin{equation}
  x^a = \overrightarrow{OX}^a,\quad x^a x_a = - 2 N^2 \Delta t^2,\quad \omega_a
  x^a = s, \quad x_a = \alpha \omega_a.
\end{equation}
The last equation expresses the fact that the tangent plane at $X$ to
$E$ in $\cS_0$ is parallel to the intersection of $H$ with
$\cS_0$. From these equations, one can easily derive the relations
\begin{gather}
  2 N^2 \Delta t^2 \omega_a \omega^a = -s^2,\\
  x^a = - \frac{2}{s} N^2 \Delta t^2 \omega^a.
\end{gather}
We are interested in coordinate planes within $\cS_0$. These are
obtained from $\omega = dx^i - N T^i dt$. In particular, we consider
coordinate planes which are a distance $\pm\Delta x^i$ from the point
$Q$. These satisfy the equation $\omega_a x^a = - N T^i \Delta t \pm
\Delta x^i$. Inserting this value for $s$ in the equations above we
obtain the equation
\begin{equation}
  \left( T^i \pm \sqrt{-2 \omega_a \omega^a} \right) N \Delta t = \pm
  \Delta x^i,
\end{equation}
which holds whenever the past light cone of $P$ touches a coordinate
plane which is $\pm \Delta x^i$ from the point $Q$. Thus, according to
our criterion, a valid time-step $\Delta t$ should satisfy
\begin{equation}
  \Delta t \le \mathrm{min}\frac{\Delta x^i}{N|T^i \pm \sqrt{-2
      \omega_a \omega^a}|},
\end{equation}
where the minimum is taken over all points in $\cS_0$.
There are some points to be mentioned:
\begin{itemize}
\item In our present case the square root is simply $\sqrt{-2
    \omega_a \omega^a}= 2\sqrt{C_{AB}^i C^{ABi}}$, so that
  determining the maximal time-step is rather simple. It involves
  going through the grid and finding the maximum of some algebraic
  function of the fields (no inversion of the spatial metric).
\item We find that the criterion is at least necessary for stability,
  i.e., if we 
  do not enforce the time-step to be at most the above value then the
  scheme becomes unstable. So far it has also been sufficient for
  stability. 
\item This criterion might be conservative. In fact, one could
  imagine that one should be able to increase the time step until the
  first of the adjacent grid points comes to lie on the null cone, the
  others still being not inside. We have not investigated this further.
\item This is a first order criterion and it might be too crude for
  the Lax-Wendroff method. One could think of enforcing this condition
  at each half step. Again this has not been investigated.
\end{itemize}

A complete time step is performed by going through the following steps
given the solution at the time level $t_n$
\begin{itemize}
\item Find the maximal possible time step $\Delta t$ by inspection of
  the current time level.
\item Set the gauge functions, also possibly according to the
  properties of the   solution at the current time level.
\item Update the solution at the interior points using the gauge
  functions and the time-step by performing the above four steps for
  each function. After the first half step specify the gauge source
  functions again.
\item Update the solution at the boundary points.
\end{itemize}

\section{Boundary treatment}
\label{sec:bndry}

Analytically, the hyperboloidal initial value problem does not need
any boundary conditions. The initial data are given on a three
dimensional manifold $\cS$ with boundary $\del\cS$ on which the
conformal factor $\Omega$ is supposed to vanish. Then a solution
exists on the four dimensional manifold $M = \cS \times
\left[0,\tau \right]$ for some $\tau > 0$ which is such that the
boundary $\del M = \del\cS \times \left[0,\tau\right]$ is a null
hypersurface and hence characteristic. That means that even if one
would extend the initial data across the boundary $\Omega = 0$ in some
way this extension could not influence the interior, i.e., the
physical space-time depends only on the data given inside the
boundary.

The situation is different in the numerical case. The characteristic
speeds are different for different modes which are propagated by the
numerical scheme. In particular, non-physical modes tend to propagate
at speeds much higher than physical propagation speeds and thus
contaminate the solution all over the entire computational domain. A
notorious place where non-physical modes are generated is at the
boundaries of the domain. Due to the lack of enough grid points there,
in general, the numerical evolution scheme has to be changed. It is
absolutely vital to impose boundary conditions so that the
non-physical modes are kept small. The GKS theory
\cite{Kreiss-1968,GustafssonKreissSundstroem-1972} which has been
developed for analysing such situations is inherently difficult to
apply. A different (equivalent) formulation based on the notion of
group velocity for finite difference schemes has been given by
Trefethen \cite{Trefethen-1982}. It has been found that certain
intuitive numerical boundary conditions do not perform as
expected. Conditions which work for one numerical scheme do not
necessarily work for others. For linear equations in one space
dimension the mathematical analysis can completely be carried
through. It turns out that the essential criterion is a non-degeneracy
condition for a linear system of equations obtained from the
combination of the evolution scheme and the boundary condition. This
system is required to have no solutions in order to exclude the
parasitic modes. Although Trefethen's method is very physical and
intuitive it does not provide enough information in the case of higher
dimensional and/or non-linear equations. It does, however, give
valuable hints as to which conditions might have a chance to be useful
in those more general situations treated with the Lax-Wendroff scheme.

The situation is somewhat ironic in the present case. One is not at
all interested in what happens at the boundary because this is
(usually) outside the physical space-time. However, it is the boundary
which needs the most careful treatment. One would wish to find gauge
conditions which make the non-physical portion of the computational
domain small, ideally putting the boundary at $\Omega=0$. We will
examine the feasibility of this idea later on.

In another aspect, the present situation is also quite
disadvantageous. Usually, it is of great importance that the
analytical problem has a well posed initial-boundary value
problem. The rigorous analysis provides the information about which
data can be specified freely at the boundary and which data is
determined from information propagating towards the boundary from the
inside. Knowledge of this kind is necessary in order not to
over-specify the solution at the boundary, because this would
inevitably lead to instabilities (except for extremely simple
cases). In our case, it is not known at present whether the system
admits a well posed initial-boundary value problem (see, however
\cite{FriedrichNagy-1997}). To overcome this lack of information we
analyse the system to first order at the boundary in the following
sense.

The boundary $\cB$ is a time-like three-dimensional hypersurface in the
space-time. Let $n_a$ be the space-like conormal of $\cB$. A system of
partial differential equations which has the form
\begin{equation}
  \del_t u_\alpha + A^i_{\alpha\beta} \del_i u^\beta = b_\alpha
\end{equation}
can be rewritten in the form
\begin{equation}
  \del_t u_\alpha + C_{\alpha\beta} \D u^\beta + 
  B^A_{\alpha\beta} \del_A u^\beta = b_\alpha,
\end{equation}
where $\D$ is the derivative along any vector field $u^a$ which
satisfies $n_a u^a=1$ on $\cB$ (usually taken to be the normal vector
field extended off $\cB$ in an arbitrary way) and where the $\del_A$
are derivatives intrinsic to $\cB$ at points of $\cB$. On the boundary
the matrix $C_{\alpha\beta}$ regulates to first order the propagation
of the fields across the boundary. By analysing its structure one can
gain valuable insights into the behaviour of the solution on $\cB$. In
particular, finding the eigenvalues and eigenvectors of
$C_{\alpha\beta}$ (which in our case is hermitian) enables us to
select combinations of the fields which (to first order) propagate
purely inward or purely outward or which stay on the boundary. These
have to be treated differently. While the ingoing pieces can be
prescribed freely, the outgoing ones have to be obtained from the
interior. This is done here by extrapolation. That this might be
possible is indicated be the Trefethen analysis which shows that
extrapolation remains stable when used in conjunction with the
one-dimensional Lax-Wendroff method. We want to mention that this
analysis applies not only at the boundary but also at the interfaces
between grid cells. This is important for possible future application
of high resolution methods which require the solution of Riemann
problems at each grid cell, see e.g., \cite{LeVeque-1997}.

To be somewhat more precise we need to analyse the three subsystems of
the full system which do not consist entirely of advection equations
along the $\del_t$ vector. These are the systems for the variables
$\left(K_{ABCD}, K_{AB}, K\right)$, the variables $\left(\phi_{ABCD},
\phi_{AB}, \phi\right )$ and for the Weyl curvature $\left(E_{ABCD},
B_{ABCD}\right)$, respectively. Note, that this analysis is valid in
the three-dimensional case. Only in the code we have specialized this
to the two-dimensional case. Let us describe the procedure for the
$\phi$-system.

First, we note that $n_a=n_{AB}$ can be viewed as a complex metric on
spin space which reduces the structure group from $SL(2,\CC)$ down to
$U(1)$. Thus, it is possible to decompose any symmetric spinors
$\Phi_{AB}$, $\Phi_{ABCD}$ into irreducible pieces with respect to the
smaller structure group:
\begin{align}
  \Phi_{AB} &= \Phi_{AB}^{(0)} - \frac{1}{2n^2} n_{AB} \Phi^{(1)}, \\
  \Phi_{ABCD} &= \Phi_{ABCD}^{(0)} - \frac{1}{n^2} n_{(AB}
  \Phi_{CD)}^{(1)} + \frac{3}{8n^4} n_{(AB} n_{CD)} \Phi^{(2)},  
\end{align}
where $n_{AB}n^{AB}=-2n^2$ and 
\begin{align}
   \Phi^{(1)} &= \Phi_{AB} n^{AB},\\
   \Phi^{(2)} &= \Phi_{ABCD} n^{AB} n^{CD},\\
   \Phi^{(1)}_{AB} &= \Phi_{ABCD} n^{CD} + \frac{1}{2n^2} n_{CD} \Phi^{(2)}.
\end{align}
This decomposition which is very natural algebraically
corresponds geometrically to a decomposition of the fields into parts
which are vertical and tangential to $\cB$. The principal part of the
$\phi$-system which does not contain tangential derivatives is
\begin{align}
  & \del \phi_{ABCD} - n_{(AB} \D\phi_{CD)},\\
  & \del \phi_{AB} - \frac23 n_{AB} \D\phi + n^{CD}\D\phi_{ABCD},\\
  & \del \phi + 2 n_{AB} \D\phi^{AB}.
\end{align}
Here we have neglected terms containing derivatives of $n_{AB}$
because those do not alter the symbol of the subsystem. Inserting the
decompositions of the variables we get the following system
\begin{align}
  & \del \phi + 2 \D \phi^{(1)}, \\
  & \del \phi^{(1)} + \frac43 n^2 \D\phi + \D \phi^{(2)},\\
  & \del \phi^{(2)} + \frac43 n^2 \D \phi^{(1)},\\
  & \del \phi_{AB}^{(0)} + \D \phi_{AB}^{(1)},\\
  & \del \phi_{AB}^{(1)} + n^2 \D \phi_{AB}^{(0)},\\
  & \del \phi_{ABCD}^{(0)}.
\end{align}
Obviously, this system can be decomposed into three smaller subsystems
and it can be shown that the coefficient matrix of the $\D$ operator
is hermitian with respect to a suitable inner product (it has to be
because it comes from a symmetric hyperbolic system). Now it is easy
to find ``characteristic combinations'' of the variables so that the
symbol becomes diagonal, i.e., it has the form 
\begin{equation}
\label{chareqn}
\del C + \lambda \D C
\end{equation}
for each characteristic quantity. These combinations are unique only
up to a scaling factor. We choose the following quantities with their
respective characteristic speeds
\begin{align}
  C_0 &= 4n^2 \phi - 6 \phi^{(2)},& \lambda &= 0 \\
  C_\pm &= 4n^2 \phi \pm 3 \sqrt{4n^2} \phi^{(1)} + 3  \phi^{(2)},&
  \lambda &= \pm \sqrt{4n^2}   \\
  C_{AB}^\pm &= \phi_{AB}^{(1)} \pm \sqrt{n^2} \phi_{AB}^{(0)},&
  \lambda &= \pm \sqrt{n^2}   \\
  C_{ABCD} &= \phi_{ABCD}^{(0)},& \lambda &= 0.
\end{align}
In an analogous way we find the characteristic quantities for the
$K$-system
\begin{align}
  C_0 &= 3 K^{(2)} + 4 n^2 K,& \lambda &= 0 \\
  C_\pm &= 6 K^{(2)} \pm 3 \sqrt{4n^2} K^{(1)} +  4n^2 K,&
  \lambda &= \pm \sqrt{4n^2}   \\
  C_{AB}^\pm &= 2K_{AB}^{(1)} \pm \sqrt{2n^2} K_{AB}^{(0)},&
  \lambda &= \pm \sqrt{2n^2}   \\
  C_{ABCD} &= K_{ABCD}^{(0)},& \lambda &= 0.
\end{align}
The Weyl system has a completely different structure from the previous
two subsystems. Nevertheless, the analysis yields characteristic
quantities written in terms of the complex Weyl spinor $\psi_{ABCD}$
and the spinor field $\varphi_{ABCD}=n_{(A}{}^E \psi_{BCD)E}$.
\begin{align}
  C_0 &= \psi^{(0)},& \lambda &= 0 \\
  C_{AB}^\pm &= \pm \sqrt{n^2} \psi_{AB}^{(1)} - \varphi_{AB}^{(1)},&
  \lambda &= \pm \sqrt{n^2}   \\
  C_{ABCD}^\pm &= \pm \sqrt{n^2} \psi_{AB}^{(0)} 
  - 2\varphi_{AB}^{(0)},& \lambda &= \pm \sqrt{4n^2}.
\end{align}
In our case, the boundary is given as a surface $v=\textrm{const}$ so
that we can put $n_{AB}=C_{AB}^2$. Let us now focus on
\eqref{chareqn}. Inserting the explicit expressions for the
derivatives this is
\begin{equation}
  \del_t C + N \left(-T^2 \del_v C + \lambda \del_v C \right) 
\end{equation}
Therefore, it is the sign of $\left( \lambda - T^2 \right)$ which
regulates in which direction the quantity $C$ propagates across the
boundary.

To update the values at the boundary points we proceed as
follows. First we determine the characteristic quantities on the
boundary. This is done by looking at the sign of the corresponding
eigenvalues which decides whether to simply set the value arbitrarily
in case the quantity propagates inwards or else whether to find the
value by extrapolation from the interior. From the characteristic
quantities we obtain the field values by reversing the transformations
above. 

In the situations considered this procedure yields a stable
algorithm. This is consistent with the Trefethen theory which shows
that in the one-dimensional case extrapolation together with the
Lax-Wendroff time evolution scheme remains stable. By its very nature
our procedure is a first order approximation to the real situation so
that we cannot expect to obtain a code which is second order accurate
in the neighbourhood of the boundary. However, since the surface
$\Omega=0$ is a characteristic surface we may hope, that the error
does not too severely influence the physical space-time as long as the
boundary of the computational domain is outside the physical space-time.

\section{Gauge choices}
\label{sec:gauge}
In this section we want to present some results about the various
possible gauge choices. Our emphasis will be on the temporal gauge
choices. There is one class of choices for the shift vector which is
natural in the present context of the conformal field equations. The
gauge for the frame rotations and the third class of gauges, namely
the choice of the scalar curvature $\Lambda$, is unknown territory as
of yet and we put the corresponding gauge source functions equal to
zero in the code. As was pointed out in I, in the case of the frame
rotations this means that the frame is Fermi-Walker transported along
the normal vector of the time foliation.

\subsection{Choices of lapse}
\label{sec:lapse}

Let us start with the temporal gauges. Fixing the lapse function is a
difficult task. This function has to be chosen in such a way that the
time coordinate does not degenerate in the course of the
evolution. Here is an attempt to collect some criteria which should be
satisfied by the lapse:
\begin{itemize}
\item the lapse function should not ``collapse'' in the sense that it
  approaches zero in a finite coordinate time,
\item the surfaces of constant time should remain smooth,
\item the lines of constant spatial coordinates should not intersect,
\item the lapse function should remain positive,
\item it should not develop too steep gradients,
\item depending on the problem the foliation should or should not avoid
  singularities.
\end{itemize}
In our treatment of the hyperboloidal initial value problem, the lapse
function cannot be chosen directly. Instead it is governed by an
evolution equation which contains the ``harmonicity'' $F=2\Box t$ as
an arbitrary function of the coordinates.  It is not easy to find a
function $F$ which allows the lapse to satisfy some or all of the
above criteria. The reason is that one has no idea what $F$ should
look like in coordinates which are constructed as the code moves
along. To make the lapse satisfy the criteria above one needs a
certain amount of ``feedback'' i.e., information about the current
status of the evolution seems to be unavoidable. This means, that one
should specify $F$ as a function of the field variables. But since in
the system also the derivatives of $F$ appear this leads to the
problem that the characteristics of the system change because the
symbol has been altered. We will discuss this later in this section.

The various choices for $N$ that have been considered so far are
\begin{itemize}
\item the ``natural gauge'' for the exact solutions obtained by
  setting $F$ equal to the expression computed from the explicit form of
  the metric (cf. appendix \ref{sec:exsol}) and variations thereof,
\item the ``Gau\ss\ gauge'' which is the condition that $N$ should be
  constant throughout the timeslice,
\item the harmonic gauge with $F=0$ and
\item the special class of gauges for which $F$ is a function of $N$
  and $K$ only, $F=f(N,K)$, which in fact includes all of the above
  gauges.
\end{itemize}
The popular ``maximal gauge'' where one requires $K=0$ probably cannot
be achieved by specifying $F$. 

We can judge the effects of these gauges by monitoring the function
$\tau$ which satisfies the eiconal equation $\nabla_a\tau \nabla^a\tau
= 1$ and the condition $\tau=0$ on the initial surface. The value of
this function at a point $P$ gives the distance in proper time between
$P$ and the intersection point of the geodesic through $P$ tangent to
the unit normal of the foliation and the initial surface. This
function is evolved simultaneously with the other field variables.
For the A3 solution the proper time distance from a point on the
initial surface $t=t_0$ with $z=0$ and the singularity at $t=z=0$ is
$\tau=2\sqrt{-t_0}$ for $t_0\le0$ and we can compare how far the
evolution proceeds with the various gauge choices. In all the examples
presented in this section a $100\times 100$ grid was used and the
coordinate system is one discussed in section \ref{sec:num} with
$v_0=5$. The initial data were taken from the exact A3 solution at an
initial time $t_0=-5$. The boundary values were specified depending on
the cases considered. When the ``natural gauge'' was used the ingoing
boundary data were taken from the A3 solution. In all other cases
these values were used initially to satisfy the corner conditions and
then specified to decrease exponentially, so that after approximately
20 timesteps the ingoing values are zero.

\subsubsection{$F$ as a function of the space-time coordinates}
\label{sec:f(t,z)}

For the ``natural gauge'' we find that we can in principle approach
the singularity arbitrarily closely by increasing the resolution
appropriately. Obviously, there are hard limits imposed on the
calculation by the finite precision arithmetic so that eventually the
numbers will overflow.

\begin{figure}[htbp]
  \begin{center}
    \leavevmode
    \makebox[\hsize]{
    \includegraphics[width=170pt]{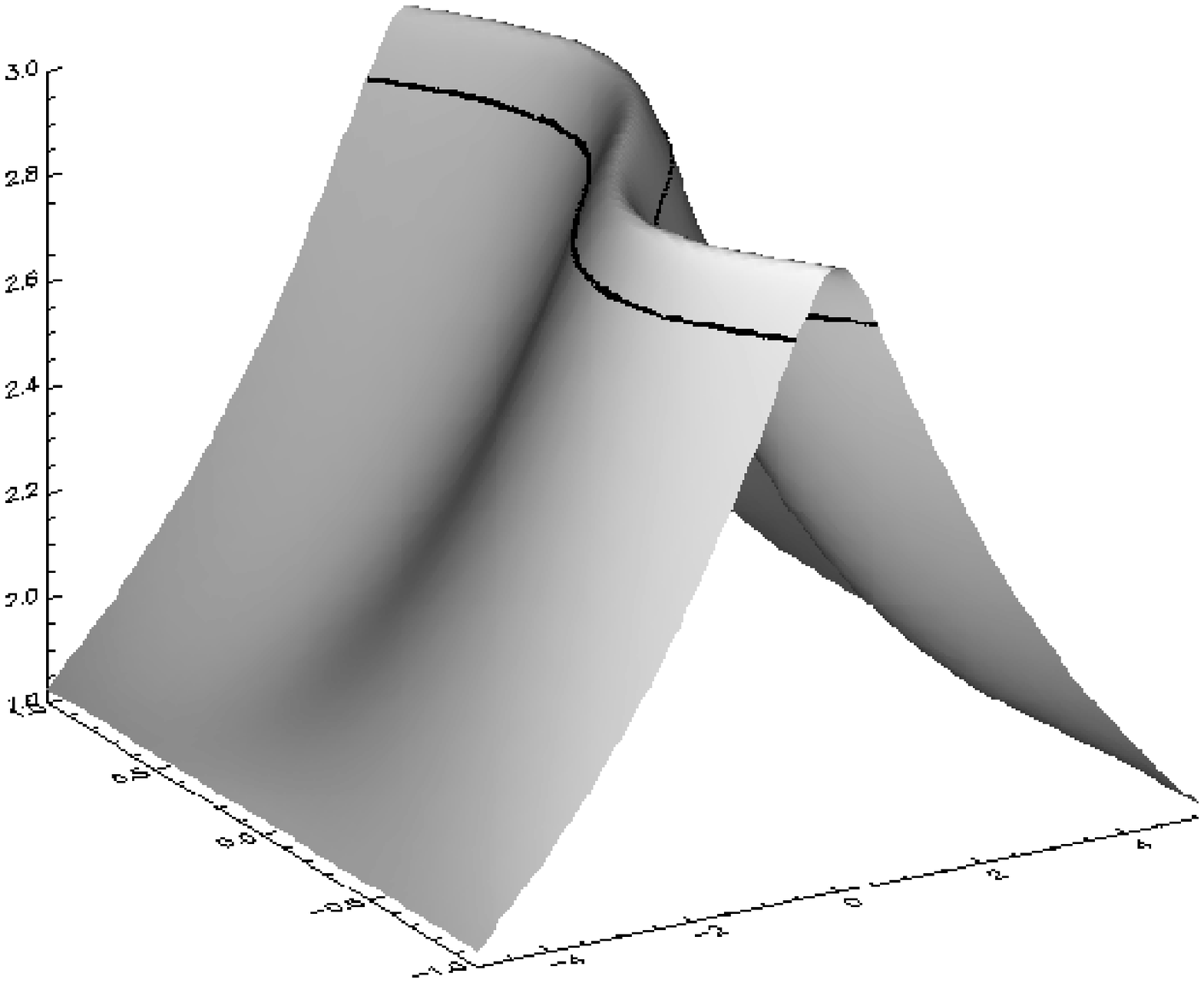}
    \includegraphics[width=170pt]{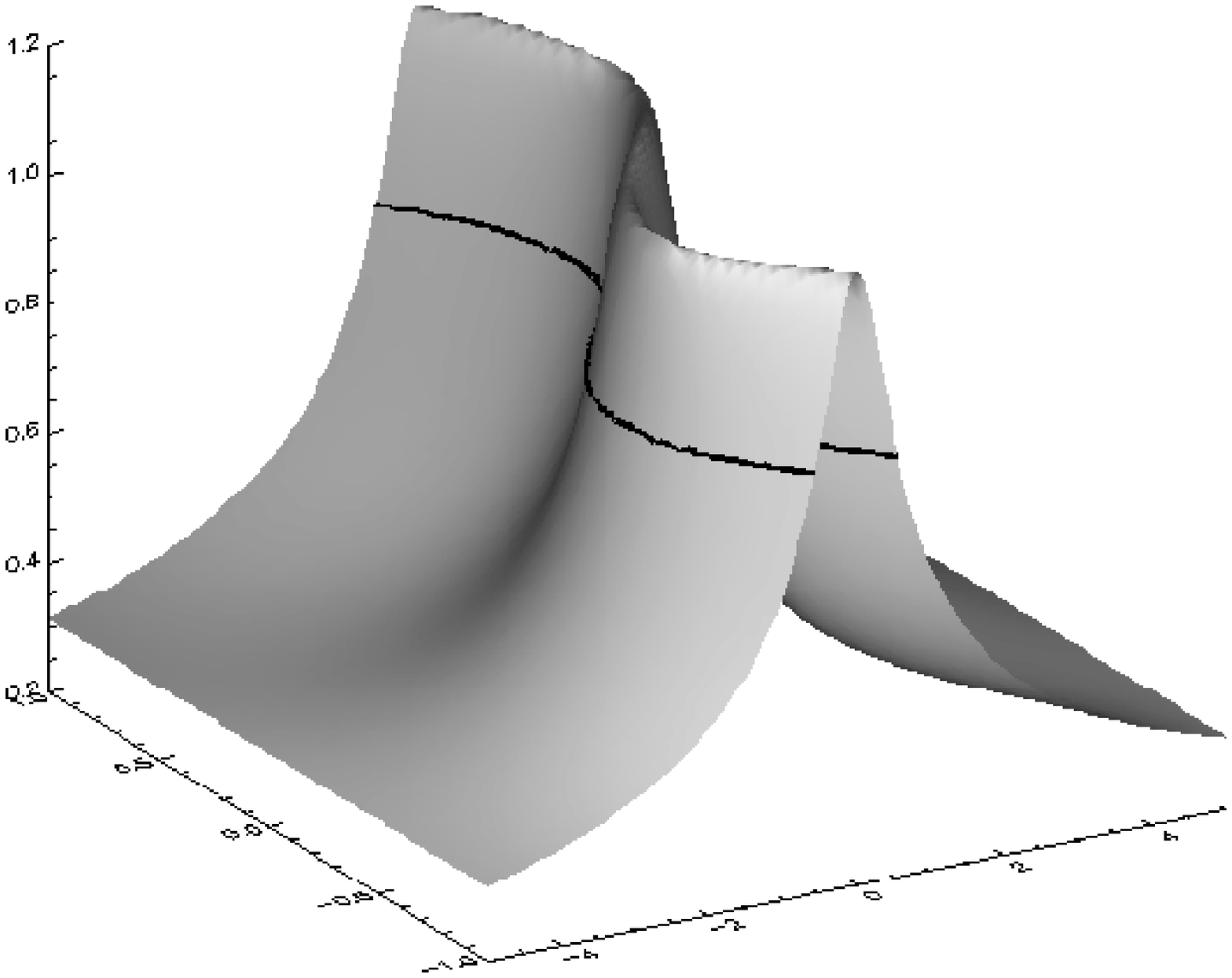}}   
    \caption{The proper time $\tau$ (left) and the lapse $N$ (right)
      in the ``natural'' gauge. The extremal values are
      $\tau_{\mbox{min}}=1.831$, $\tau_{\mbox{max}}=2.988$ and
      $N_{\mbox{min}}=0.315$, $N_{\mbox{max}}=1.151$. The black
      contours show the locations of the two $\scri$'s.} 
    \label{fig:tau_nat}
  \end{center}
\end{figure}

\begin{figure}[htbp]
  \begin{center}
    \leavevmode
    \makebox[\hsize]{
    \includegraphics[width=170pt]{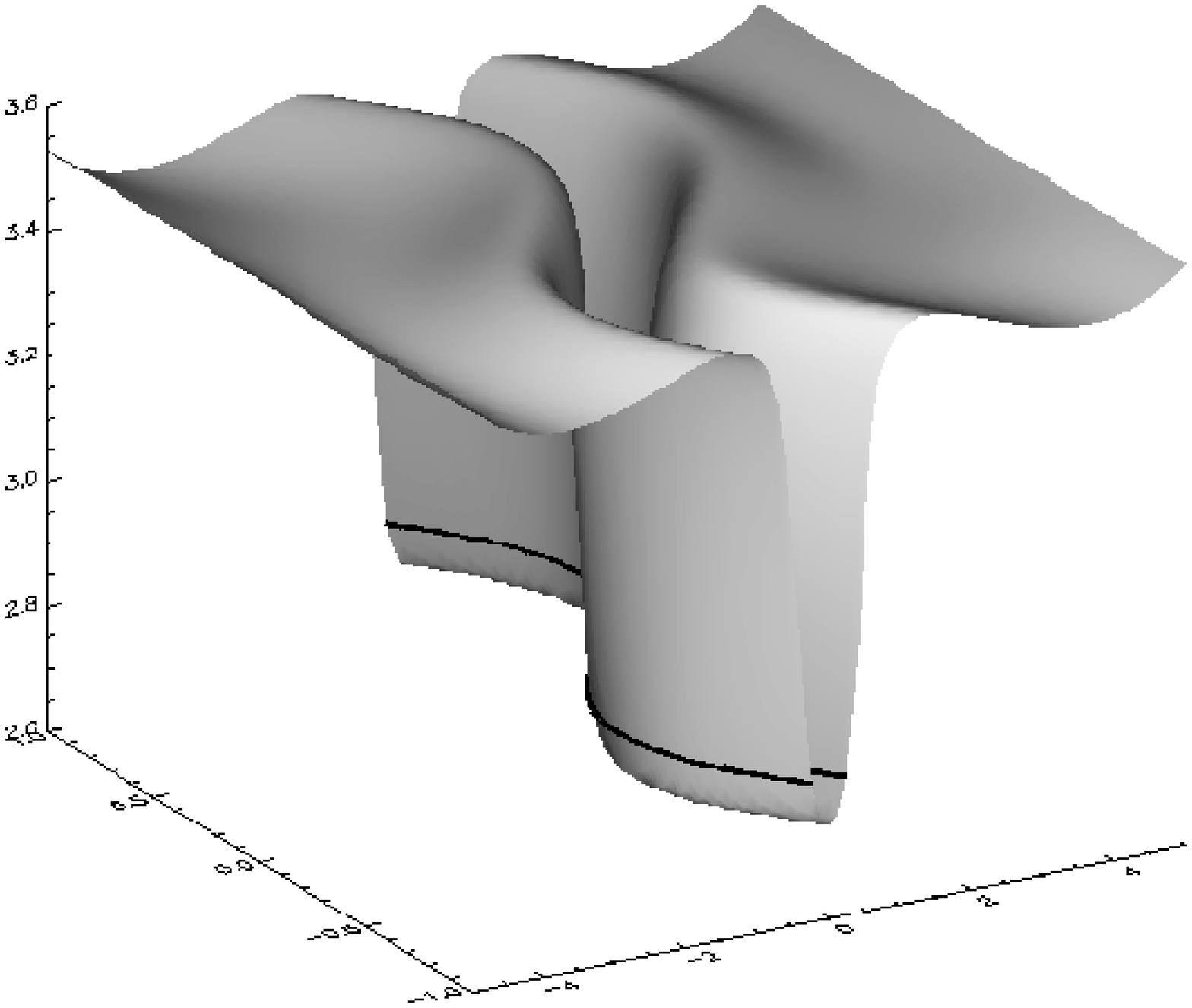}
    \includegraphics[width=170pt]{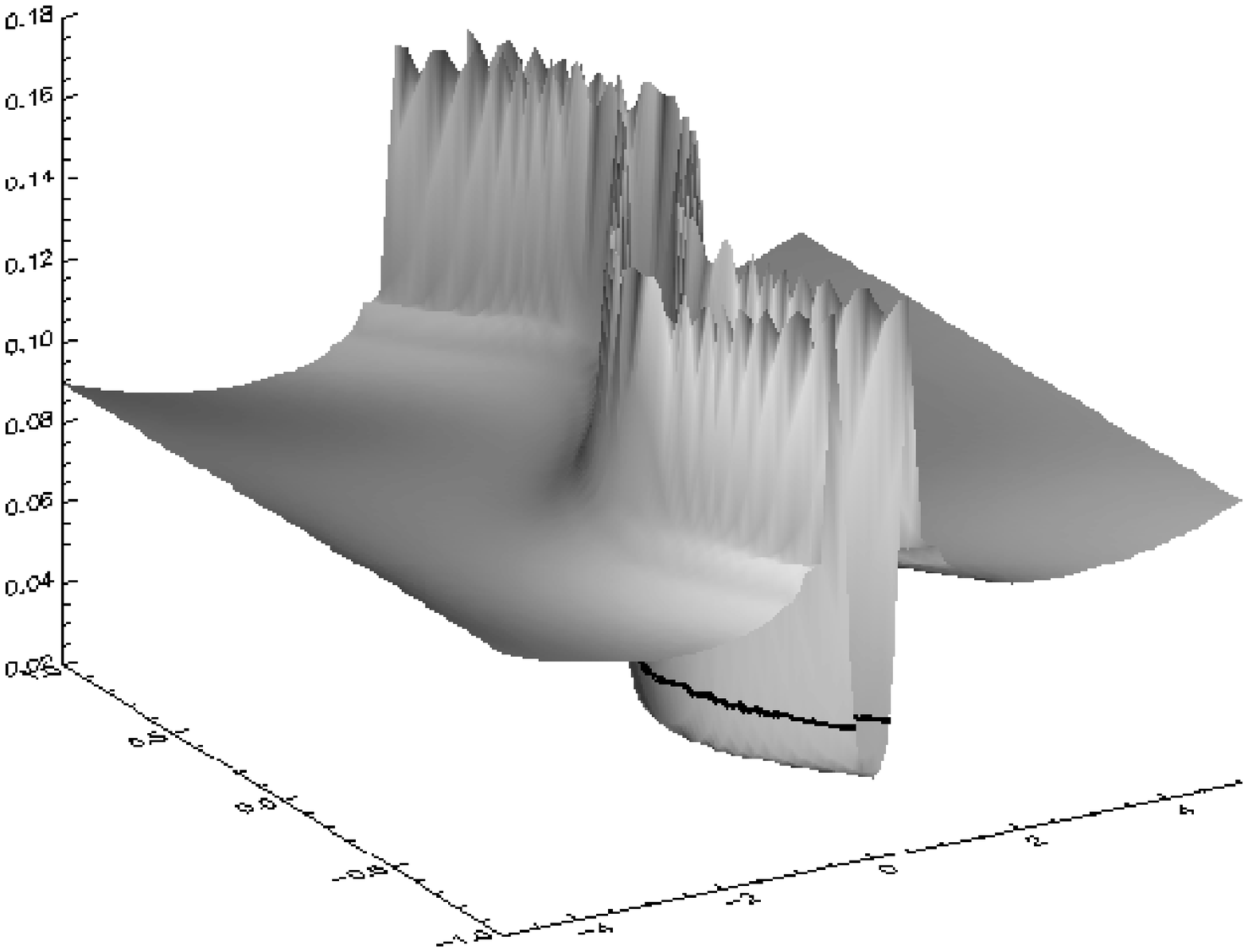}}   
    \caption{The proper time $\tau$ (left) and the lapse $N$ (right)
      in the harmonic gauge. The extremal values are
      $\tau_{\mbox{min}}=2.747$, $\tau_{\mbox{max}}=3.537$ and
      $N_{\mbox{min}}=0.03964$, $N_{\mbox{max}}=0.1624$. The zig-zag
      behaviour in the figure for $N$ is due to a lack of sufficient
      resolution for the shading process.} 
   \label{fig:tau_harm}
  \end{center}
\end{figure}

The difference between the natural and the harmonic gauge is shown in
Fig.~\ref{fig:tau_nat} and Fig.~\ref{fig:tau_harm}. In both figures we
show the proper time $\tau$ and the lapse at a late instant of the
time evolution. For the natural gauge this was dictated by the code
which stopped because the time step could not be chosen without
violating the CFL condition. This is due to the closeness of the
singularity. The light cones are infinitely stretched along the
symmetry directions as the singularity is approached. A better
resolution would have extended the evolution time somewhat more but
eventually the same thing would happen. In this run the coordinate
time elapsed was roughly 4.405. It is clear that with this temporal
gauge the singularity can be reached (at least in principle).

With harmonic gauge the code was stopped after an elapsed coordinate
time 20.07. In Fig.~\ref{fig:tau_harm} it can be seen that the
evolution close to the singularity has slowed down considerably. The
proper time in the center is much smaller than for regions further
outside. The lapse function is very small in the center and it is
decreasing. It is not known whether the evolution will reach the
singularity even in principle. The decrease in the lapse could be so
rapid that the integrated proper time along the central geodesic would
reach a limit below the value $2\sqrt{-t_0}$. It is quite likely, that
in this gauge the code will ultimately not be able to resolve the
steep gradients which occur at late times between the interior region
which cannot progress beyond the singularity and the exterior region
which can. 

Related to this phenomenon is the fact that the interior region
shrinks. On the initial surface $\scri$ is located at the boundary of
the grid and it gradually moves inward during the
evolution. Ultimately, there will be only very few grid points left in
the interior region. This is a phenomenon which has nothing to do with
the temporal gauge but with the choice of the shift vector. We will
discuss later in this section a shift gauge which allows the freezing
of $\scri$ on the grid.

To get some feeling for the influence of $F$ on the time slicings we
study the slicings obtained from substituting $p\cdot F$ for $F$ with
some parameter $p$. For $p=0$ we have the harmonic gauge which avoids
the singularity. For $p=1$ we have the natural gauge which allows us
to reach the singularity in finite time. What happens when we increase
$p$ beyond unity?

\begin{figure}
  \begin{center}
    \leavevmode
    \hspace{1cm}\includegraphics[height=6cm]{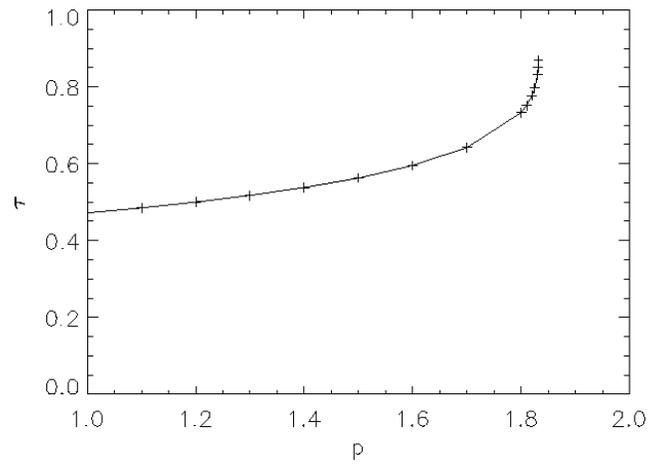}
    \caption{ Proper time $\tau$ as a function of the
      parameter $p$}
    \label{fig:tau_p}
  \end{center}
\end{figure}

\begin{figure}
  \begin{center}
    \leavevmode
    \includegraphics[height=6cm]{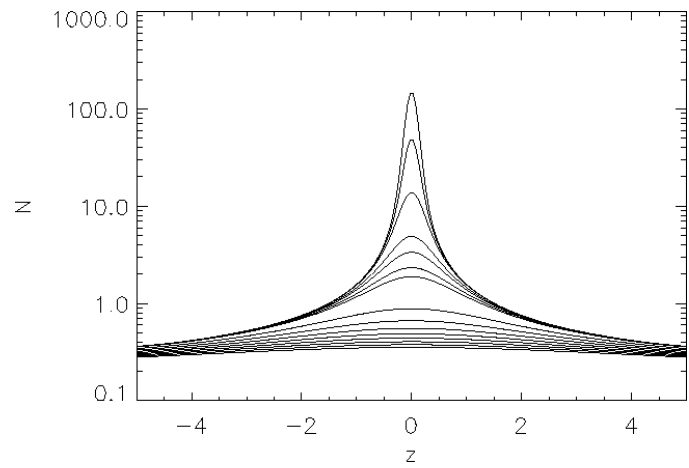}
    \caption{Profiles of the lapse N for a fixed value of the Killing
      coordinate $x$ and different values of $p$. Note the logarithmic
      scale.}
    \label{fig:N_z}
  \end{center}
\end{figure}

\begin{figure}
  \begin{center}
    \leavevmode
    \includegraphics[height=6cm]{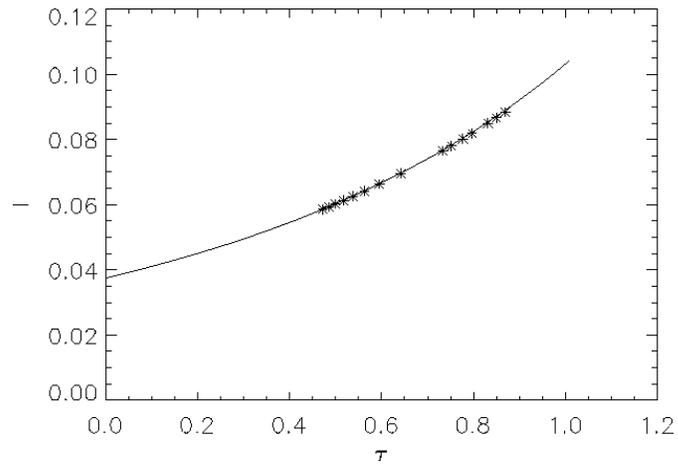}    
    \caption{The curvature invariant $I$ along $z=0$ as a function of
      proper time. The line is the exact function, the points indicate
      computed values.} 
    \label{fig:I_tau}
  \end{center}
\end{figure}

We evolve for a fixed coordinate time interval $t\in [-5,-4]$ with
different values of $p\in \left[1.0, 1.8315 \right]$. We find
that for $p>1.8315$ the code crashes. It is easily seen that this
crash cannot be due to the curvature singularity in the A3 solution
but is a coordinate singularity. In Fig.~\ref{fig:tau_p} we plot the
proper time distance between the initial and final time slice along
the central geodesic ($z=0$) versus the parameter $p$. We see that
$\tau(p)$ has an infinite derivative at $p=1.8315$ with a finite value
of $\tau$ far from its value $2\sqrt5$ at the singularity.
The lapse function $N$ for different values of $p$ diverges rapidly
(cp. Fig. \ref{fig:N_z}). The curvature invariant
$I=\Omega^2\left(E_0E_4+6E_2^2\right)$ which diverges at the
singularity stays perfectly regular. Fig. \ref{fig:I_tau}
shows the exact invariant plotted against proper time along the
central geodesic. The dots are the values of $I$ and $\tau$ obtained
from the runs with different parameter values. We see that the
behaviour of these functions is not altered by the occurrence of the
coordinate singularity.

\begin{figure}
  \begin{center}
    \leavevmode
    \includegraphics[height=6cm]{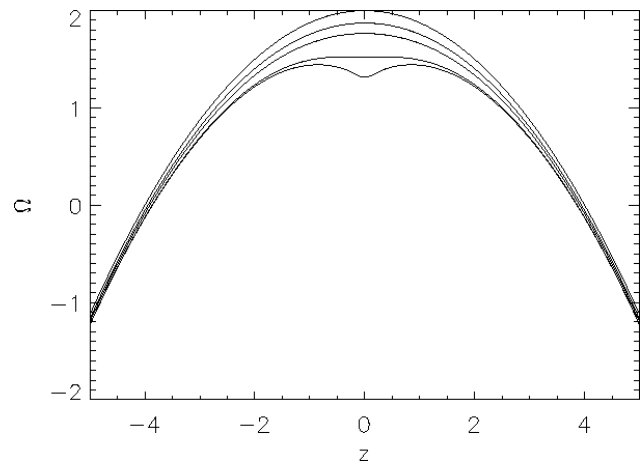} 
    \caption{Profiles of the conformal factor $\Omega$ for some values
      of the parameter $p$ } 
    \label{fig:Om_z}
  \end{center}
\end{figure}

Finally, we show in Fig. \ref{fig:Om_z} 
the profiles of the conformal factor for various values of $p$. As the
parameter approaches its final value the conformal factor develops a
minimum at the center. Although this behaviour seems strange at first
sight it can easily be explained. The conformal factor decreases as a
function of $\tau$ for fixed spatial coordinates. Due to the rapid
divergence of the lapse the proper time at $z=0$ is much larger than
in the regions outside. So that we see values of $\Omega$ in the
center which are reduced over proportion from the values outside the
center. This accounts for that central dip.

\subsubsection{The Gau\ss\ gauge}
\label{sec:gauss}

The ``Gau\ss\ gauge'' which forces $N$ to be constant is a condition
which is imposed on the lapse function directly. In principle, it is
possible to express the exact solutions in Gau\ss\ coordinates by
performing the coordinate transformation explicitly. Then one can
compute the ``harmonicity'' for these coordinates and do the
evolution. However, we proceed somewhat differently to impose the
Gau\ss\ gauge.  The lapse function satisfies the evolution equation
\cite[\eqref{eq:evN}]{jf-1997-2}
\begin{equation}
  \hat\del N = - N^2 K - N^3 F
\end{equation}
where $\hat\del=N\del$. Now suppose that $N$ would satisfy an
evolution equation 
\begin{equation}
\label{eq:Ndot}
  \hat\del N = \alpha (\bar N - N),
\end{equation}
where $\bar N$ is an arbitrary (positive) constant. This equation has
the solution $N(s)= \bar N + N_0 e^{-\alpha s}$, $s$ being the
parameter with $\hat\del s = 1$. Thus, for $\alpha > 0$ the lapse $N$
approaches the constant value $\bar N$ in the course of the
evolution. We can make $N$ satisfy the above evolution equation
\eqref{eq:Ndot} by choosing
\begin{equation}
  F = \frac{\alpha(\bar N - N) + N^2 K}{N^3}.
\end{equation}
We find what one would expect, namely that in this gauge the
timeslices develop caustics (or, what has become known as ``coordinate
shocks''). This makes the Gau\ss\ gauge inappropriate for long time
evolutions.

\subsubsection{$F$ as a function of $N$ and $K$}
\label{sec:F(N,K)}

Let us now focus on the class of gauges defined by $F=f(N,K)$. Among
this class there is a subclass for which the lapse depends only on the
three-dimensional volume (-element) $V$, $N=g(V)$. For these gauges,
we have with appropriate assumptions on the function $g$ 
\begin{equation}
  \del N = g'(V)\, \del V = -K g'(V)\, V = -K g'(g^{-1}(N))\, g^{-1}(N),
\end{equation}
so that, in fact, $F$ is a function of $N$ and $K$ only. The natural
gauge falls into this subclass with $g(x)=x^{-\frac13}$ and,
consequently, $F=-\frac{4K}{3N}$. Similarly, the harmonic gauge with
$F=0$ is in this class with $g(x)\propto x$.

If we specify $F$ for the natural gauge not as a function of the
space-time variables but instead as depending on the field variables,
then an interesting phenomenon occurs. Although nothing else in the
code has been changed it seems to notice this difference because the
boundary becomes unstable very quickly. However, inside the
computational domain we get the same solution without any significant
difference between the two ways of specifying the gauge.

This phenomenon can be traced to the fact mentioned above, namely that
the gauge specification might change the characteristics of the
system. We can see this explicitly as follows. In equation
\cite[\eqref{eq:evK2}]{jf-1997-2} the derivative of $F$ appears. With
$F=f(N,K)$ the principal part of that equation is
\begin{equation}
    \del K_{AB} + 2 \del^{CD} K_{ABCD} + 2N f_{,K} \del_{AB} K 
    + 2N f_{,N} \del_{AB} N .
\end{equation}
The term involving $\del_{AB} N$ can be removed by using the
constraint equation \cite[\eqref{eq:cnN}]{jf-1997-2} so that the symbol for the
$K$-subsystem is $ (p,p_{AB}) \mapsto S_p(L,K)$, where $S_p$ denotes
the sesquilinear form obtained from the principal part of the
$K$-system by replacing the derivative operators $(\del,\del_{AB})$
with $(p,p_{AB})$ and multiplying appropriately with the complex
conjugate of some spinor fields  $(L_{AB}, L_{ABCD})$. Thus,
\begin{multline}
  S_p(L,K) = \frac{p}{2} \;\widehat{L}^{AB} K_{AB} 
  + \widehat{L}^{AB}p^{CD} K_{ABCD} 
  + (N f_{,K})  \widehat{L}^{AB}p_{AB}\; K\\
  + p\widehat{L}^{ABCD} K_{ABCD} - \widehat{L}^{ABCD}p_{AB} K_{CD}.
\end{multline}
Various important properties of the $K$-system can be determined from
the form $S_p$. In particular, the system is symmetric if $S_p$ is
hermitian, i.e., if $S_p(L,K) = \overline{S_p(K,L)}$ for all
$(p,p_{AB})$. It will be symmetric hyperbolic iff there exists
$(p,p_{AB})$ such that $S_p$ is positive definite. We see from the
above that the $K$-system can be made symmetric if we do not consider
equation \cite[\eqref{eq:evK4}]{jf-1997-2} but add to that equation an
appropriate multiple of its trace. This changes $S_p$ into (with
$L=L^{AB}{}_{AB}$)
\begin{multline}
  S_p(L,K) = \frac{p}{2} \;\widehat{L}^{AB} K_{AB} 
  + \widehat{L}^{AB} p^{CD} K_{ABCD} 
  + (N f_{,K})  \widehat{L}^{AB}p_{AB}\; K\\
  + p\; \widehat{L}^{ABCD} K_{ABCD} -
  \widehat{L}^{ABCD}p_{AB} K_{CD}   \\
  + (N f_{,K}) p\; \widehat{L} K -
  (N f_{,K}) \widehat{L}\; p^{CD} K_{CD}.   
\end{multline}
It is easy to see that this form is hermitian and that it will also be
positive definite provided that the inequality 
\begin{equation}
\label{eq:ineq}
  \alpha \equiv Nf_K + 1/3 > 0
\end{equation}
is satisfied. This is of course a restriction on the possible gauges.

Furthermore, the characteristics of the above system can be obtained
by inspection of its characteristic polynomial defined by
\begin{equation}
  P(p,p_{AB}) = \det(S_p),
\end{equation}
for which we obtain the expression
\begin{equation}
\label{charpol}
  P(p,p_{AB}) = 24 \alpha p^3 \left(p^2 + p^{AB} p_{AB} \right)^2
  \left(p^2 + 2 \left(\alpha + \frac23\right) p^{AB} p_{AB} \right).
\end{equation}
This polynomial is homogeneous of degree nine in its variables and,
regarded as a polynomial in $p$ only, it will have nine real zeroes
provided that $\alpha + \frac23$ is positive, which is always the case
if the inequality \eqref{eq:ineq} is satisfied. In this case, there
will exist three different characteristics, namely the lines along the
time evolution vector, the cone given by the first factor in
parenthesis in \eqref{charpol} which is double layered and a simply
layered cone given by the last factor in \eqref{charpol}. The latter
cone is gauge dependent while the former is not. The degenerate
characteristic is time-like while the gauge dependent characteristic
has no gauge independent causal character. The cases when $F$
vanishes, the harmonic gauge, or when $F$ is specified as a space-time
function correspond to $\alpha=1/3$ in which case the gauge dependent
characteristic coincides with the light cone. For the other cases $0 <
\alpha < 1/3$ and $1/3 < \alpha$ the characteristic is time-like,
respectively space-like. However, there are gauges within the
specified class for which this characteristic does not even
exist. Thus, the system acquires a mixed type, having hyperbolic and
elliptic parts. In particular, for the natural gauge specified in
terms of field variables we have $Nf_{,K}+1/3=-1$ which violates the
inequality \eqref{eq:ineq}.

A natural question to ask is the following: to what extent are these
features noticeable in the code? Judging from experiments what seems
to be the case is that the code will probably not detect differences
in the various cases as long as it does not make use of the hyperbolic
character of the system. In particular, it will probably not detect
when the system changes its character from a hyperbolic type to a
mixed type due to a gauge change. However, in those instances where
the hyperbolic character is in fact used in the code difficulties will
arise. In the present code we find that the boundary will become
unstable very quickly when we choose a gauge which makes the system
partly elliptic (we use this term only to indicate that the resulting
system is no longer hyperbolic). This is of course due to our
treatment of the boundary which implicitly assumes that $F$ is
specified as a coordinate function. Another instance which can detect
gauge changes is due to the time-step control. Here, we implicitly
assume that the largest propagation speed is the speed of light. For
gauges with $\alpha > 1/3$ this is not the case, the largest
propagation speed is bigger than the speed of light. But the largest
speed is the one which limits the time-step in order to enforce the
Courant-Friedrichs-Lewy condition for stability of the code. And, in
fact, choosing $\alpha$ big enough results in numerical instabilities
inside the computational domain.

These tests have been performed using the initial data of the A3
solution and then specifying various gauges by choosing $F$. We find
the surprising feature indicated already above that the code detects
whether $F$ is specified as a space-time function
\begin{equation}
  F= \frac{4t}{\sqrt{t^2+z^2}}
\end{equation}
or as a function of lapse and mean curvature
$F=-\frac43\frac{K}{N}$. While it runs without problems in the former
case all the way up to a maximum of the proper time $\tau$ close to
its theoretical limit in the latter case the boundary becomes unstable
very quickly. As surprising as this might seem it is still in
accordance with the general picture. What might be even more
surprising is the fact that in the interior there is apparently no
sign of any difference between the two cases.

Another gauge which has some geometric significance is given by
choosing $N \propto \sqrt[3]{V}$. This condition can be obtained from
the requirement that the height of the backward light cone of a point
in the next time level should be proportional to the ``volume radius''
$R=\sqrt[3]{V}$ of its intersection with the current time level. This
condition is satisfied for the standard $t$-foliation in Minkowski
space. Thus, we have
\begin{equation}
  \frac{\del N}{N} = \frac13 \frac{\del V}{V} =  -\frac13 K
\end{equation}
and $F= \frac23 \frac{K}{N}$. The speed of the gauge modes is in this
case bigger than the speed of light but the system remains hyperbolic.
In practice, this gauge is not very much different from the harmonic
gauge. 

What we learn from these various discussions and experiments is that
the natural gauge is the most efficient one for approaching the
singularity. However, in situations where there is no exact solution
this gauge is not available. Now one has various possibilities: one
could prescribe a gauge condition once and for all like the ones
considered in section~\ref{sec:F(N,K)} or even like the maximal gauge
where one needs to solve an elliptic equation on each timeslice. The
former have the disadvantage that they introduce superluminal
propagation speeds into the problem so that the stability of the
(explicit) methods forces rather small time steps while the latter are
rather time consuming. The other approach would be to always specify
$F$ as a function of the coordinates. This means that one needs to
experiment in order to find a good candidate expression for $F$ which
allows to reach singularities effectively. This method is very
flexible but it is also rather obscure because there are no guiding
principles about the shape of the harmonicity function $F$.

\subsection{Choices of shift vector}
\label{sec:shift}

The choice of a shift vector is even more obscure. There are two
issues involved in the choice of the shift vector: the problem of what
to do at the points of the physical space-time and how one is to treat
the points on $\scri$. 

Let us first discuss the interior issues. To describe the problems
involved we focus on the lines of constant spatial coordinates
parametrised by the coordinate time. These are the integral curves of
the vector field $\tau=\frac\del{\del t}$, the ``$t$-lines'', which
form a family of time-like lines. It is the geometry of that
congruence which can be influenced by choosing the shift vector. To
discuss this in more detail we decompose the time-like coordinate
vector into lapse and shift
\begin{equation}
  \label{eq:5.1}
  \tau^a = N\left( t^a + T^a \right)
\end{equation}
and we choose a connecting vector $\xi^a$, i.e., a vector field which
commutes with $\tau^a$. Such a connecting vector, which is also called
a Jacobi field, can be viewed as describing an infinitesimally
separated line in the family with $\xi^a$ connecting points with the
same value of the time parameter. Thus, $\xi^a$ is tangent to the
$t=\mathrm{const}$ surfaces, satisfying $\xi^at_a=0$. From the
commutator of the two vector fields we obtain
\begin{equation}
  \label{eq:5.2}
  \tau^a \nabla_a \xi^b = \frac1N \xi^a\nabla_aN \tau^b + N
  \left(\xi^a\nabla_a t^b + \xi^a\nabla_a T^b  \right).
\end{equation}
The contraction of this equation with the time-like normal of the
surfaces yields the constraint equation
\cite[\eqref{eq:cnN}]{jf-1997-2} which couples the gradient of $N$ to
the time evolution of the acceleration vector. The other part of the
equation which is intrinsic to the hypersurfaces can be obtained by
projecting \eqref{eq:5.2} along $\tau^a$ onto the hypersurfaces. This
is achieved by contraction with the projection operator
\begin{equation}
  \label{eq:5.3}
  p^a{}_b = \delta^a{}_b - \frac1N t_b \tau^a.
\end{equation}
This yields the relation
\begin{equation}
  \label{eq:5.4}
  \dot{\xi}^a = N \left( \xi^b K_b{}^a + \xi^b \del_b T^a \right),
\end{equation}
where the dot simply means $\tau^a\nabla_a$ followed by projection. As
in the case of geodesic congruences this family of $t$-lines can be
described infinitesimally by its twist, shear and divergence according
to the irreducible decomposition of the right hand side of
\eqref{eq:5.4}. From this we can conclude that a constant shift vector
generally causes the family of $t$-lines to shear and diverge,
depending on the properties of the extrinsic curvature. This is well
known in the case of Gau\ss\ coordinates which develop conjugate
points unless the hypersurface is very special.

The goal of choosing a shift vector should be to prevent the $t$-lines
from coming too close together. The twist of the congruence,
entirely due to the shift vector, does not change the relative
distances of the $t$-lines. Therefore, we need to eliminate as many
components as is possible from the shear and divergence combined in
\begin{equation}
  \label{eq:5.5}
  \sigma_{ab} = \del_{(a} T_{b)} + N K_{ab}.
\end{equation}
Since there are only three components in the shift vector, only three
components of $\sigma_{ab}$ can be compensated. Depending on which
components are to be eliminated there result different, and in general
elliptic, equations to be satisfied by $T^a$. One possibility is to
eliminate the divergence of the congruence which leads essentially to
a Poisson equation. Another possibility to determine a shift vector is
not to eliminate components of $\sigma_{ab}$ but to minimise the
functional $\int \sigma_{ab} \sigma^{ab} dV$. This leads to the
well-known ``minimal distortion'' shift condition, which is a second
order elliptic equation for the shift vector. The problems related to
the interior of $\scri$, i.e., to the physical space-time are
essentially the same as in the numerical treatment of the traditional
Cauchy problem, and there is no insight to be gained from the
hyperboloidal initial value problem.

However, this is different when one looks at the issues concerned with
the boundary of the physical space-time. One objection against the use
of conformal methods in the numerical treatment of the Einstein
equations has been the following: as the evolution proceeds the part
of $\cS$ which corresponds to the physical space-time shrinks so that
there are less and less grid points left in the interior of $\scri$
(see the figures \ref{fig:tau_nat} and \ref{fig:tau_harm}). This
implies that the resolution of features in the physical space-time is
getting smaller. However, as it turns out, this is a misconception
which might be caused by the familiar conformal diagrams of
asymptotically flat space-times. There it is assumed that light rays
are aligned on $45^o$ lines. This need not be the case. In fact, by
choosing the shift vector on $\scri$ appropriately we gain complete
control over the movement of $\scri$ through the grid. Therefore, we
get to choose between (at least) two options. On the one hand, we can
compute a Penrose diagram of the space-time which is useful for
discussing its global properties. E.g., it helps in deciding whether
there exists a regular $i^+$ or whether there appear singularities
before $i^+$ can be reached. Another option is to have $\scri$ not
move at all through the grid. This enables one to keep the resolution
in the interior constant so that the physical space-time does not
suffer any loss of resolution during the evolution. This property is
desirable when studying the behaviour of sources in the physical
space-time. Although in this case, the picture which emerges looks
like the one obtained by spatially compactifying space-time one should
keep in mind that the conformal structures are entirely different in
the two cases. After all, in the picture proposed here, $\scri$ is
still a regular characteristic surface.

How can we achieve that $\scri$ does not move through the grid? The
equation for the conformal factor is 
\begin{equation}
  \label{eq:5.6}
  \del_t \Omega = N\left( T^i \del_i \Omega + \Sigma\right).
\end{equation}
Note, that $T^i \del_i \Omega = T^{AB} \del_{AB} \Omega = T^{AB}
\Sigma_{AB}$. Thus, if we choose
\begin{equation}
  \label{eq:5.7}
  T_{AB} = 2 \frac{\Sigma_{AB}}{\Sigma},
\end{equation}
then we obtain the equation
\begin{align}
  \label{eq:5.8}
  \del_t \Omega &= \frac{2N}{\Sigma}\left( \Sigma_{AB} \Sigma^{AB} +
    \frac12 \Sigma^2 \right) \\
                &= \frac{4\Omega N}{\Sigma}\left( S - \Omega
                  \Lambda \right).
\end{align}
Therefore, $\del_t \Omega$ is proportional to $\Omega$ so that
$\Omega$ remains zero along the $t$-lines at those places where it was
zero in the beginning of the evolution, i.e., on $\scri$. This implies
that $\scri$ does not move through the grid. Although it looks as if
the shift vector is now uniquely fixed, this is not the case. Note,
that the choice
\begin{equation}
  \label{eq:5.9}
  T_{AB} = 2 \frac{\Sigma_{AB}}{\Sigma} + \Omega \tilde T_{AB}
\end{equation}
exhibits the same behaviour. Here $\tilde T_{AB}$ is completely
arbitrary apart from the fact that it should be bounded on
$\scri$. Its only effect is on the coordinates in the interior. If we
choose $\tilde T_{AB}$ so that $\Omega \tilde T_{AB}$ has finite
values on $\scri$ then we can achieve that $\scri$ moves through the
grid in a rather arbitrary but controlled fashion.

It should also be pointed out that the form of the shift vector given
in \eqref{eq:5.9} is unique, imposed by the geometry. It does not
suffer from the shortcomings of other gauge choices. In fact,
although it is specified by prescribing $T_{AB}$ as a function of the
dependent variables, this does not change the characteristics of the
system even though there are terms involving the derivatives of the
shift vector.

\begin{figure}[htbp]
  \begin{center}
    \leavevmode
    \makebox[\hsize]{
    \includegraphics[width=170pt]{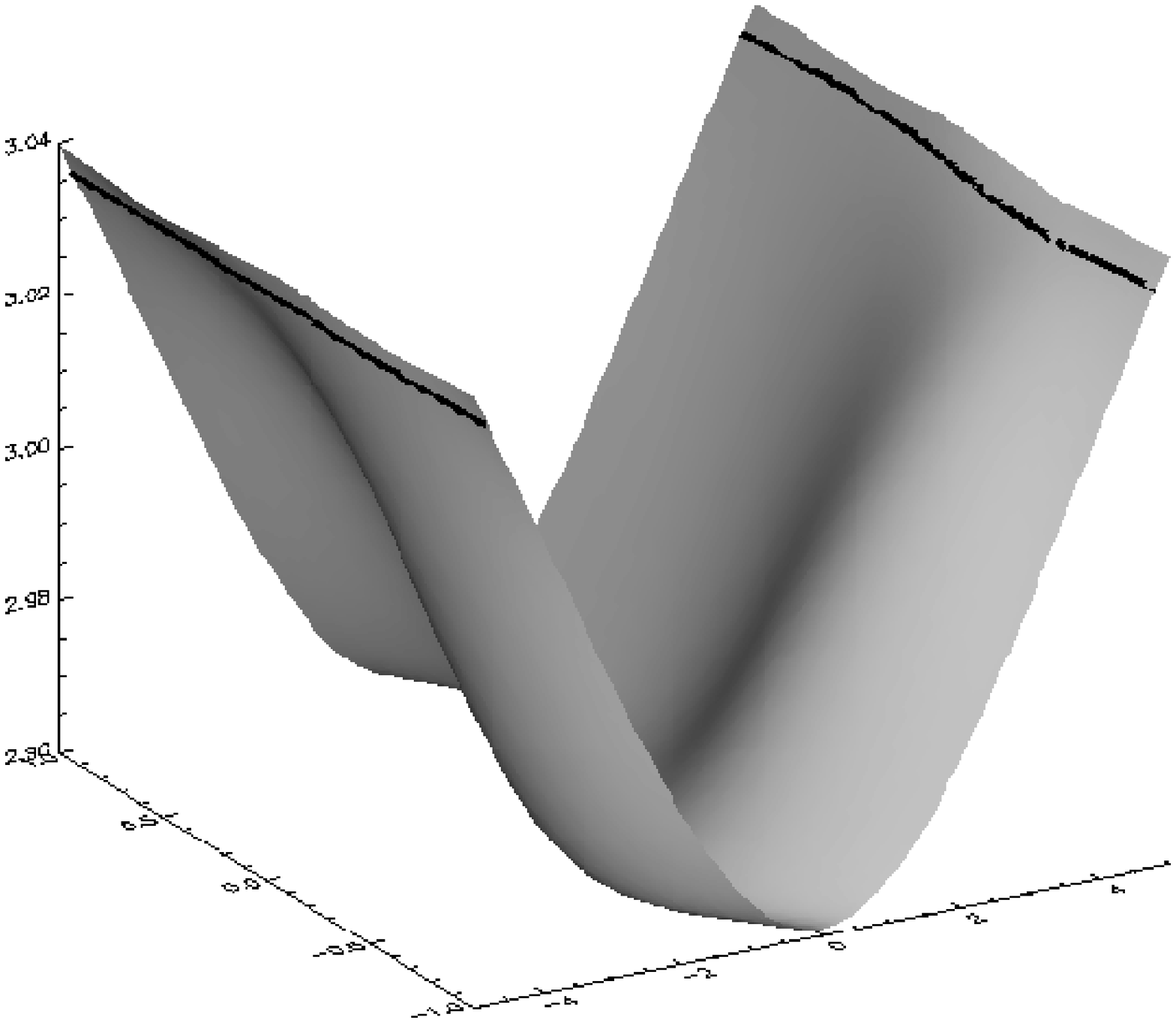}
    \includegraphics[width=170pt]{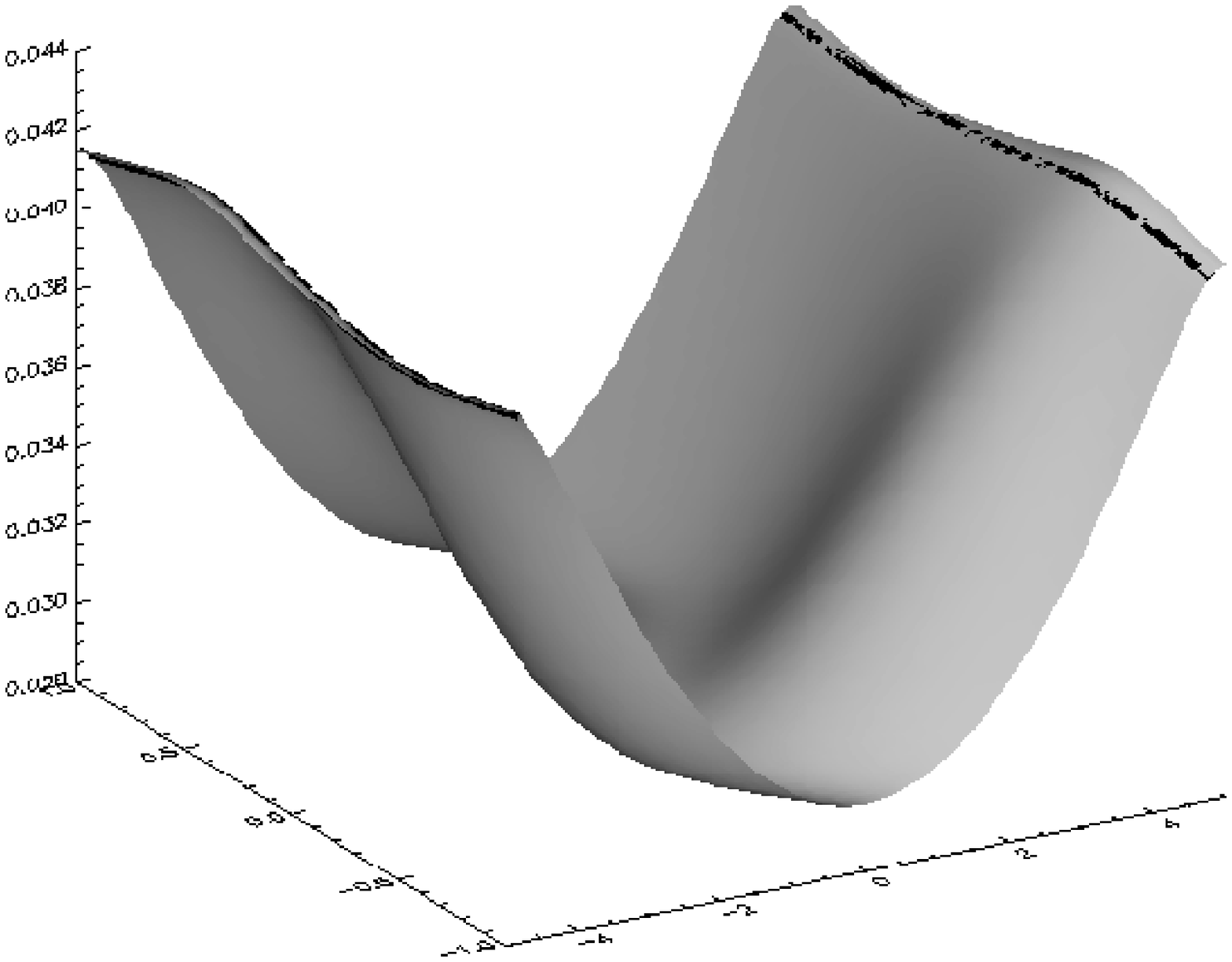}}   
    \caption{The proper time $\tau$ (left) and the lapse $N$ (right)
      in harmonic gauge with $\scri$ freezing. The extremal values are
      $\tau_{\mbox{min}}=1.831$, $\tau_{\mbox{max}}=2.988$ and
      $N_{\mbox{min}}=0.315$, $N_{\mbox{max}}=1.151$. The black
      contours show the locations of the two $\scri$'s.} 
    \label{fig:tau_harm_fr}
  \end{center}
\end{figure}

In Fig.~\ref{fig:tau_harm_fr} we show the proper time and the lapse
function for a run with harmonic gauge and scri freezing. The initial
location of $\scri$ was on the boundary $v_0 = \pm5$. The length of
coordinate time spent was $t_1-t_0=25$ with roughly 1000
time steps. $\scri$ has moved during this evolution at most over 3 grid
points. This is due to numerical inaccuracies. We see from the figures
that the evolution is much more homogeneous over the interior with
differences in proper time within the interval
$\left[2.96,3.04\right]$. But we also see that the lapse has decreased
rapidly, from a maximum value of $0.316$ at the beginning to a maximum
value of $0.0422$ in the end.

\section{Mass loss and radiation}
\label{sec:massrad}

The main motivation to consider the conformal field equations in the
first place is the claim that having $\scri$ at finite places allows a
well defined numerical description of the asymptotic properties
like the radiative information (such as shear and news on $\scri$) and
also the global properties like the Bondi energy-momentum and angular
momentum. From the nature of the hyperboloidal initial value problem
it is clear that we cannot get our hands on the ADM quantities which
are located at space-like infinity $i^0$ which is not in the domain of
dependence of any hyperboloidal initial surface.

In the numerical treatment there exists a natural foliation of $\scri$
into two-dimensional cross sections or ``cuts'' which is obtained from
the intersection of $\scri$ with the constant time hypersurfaces.  The
news, shear and the ``null datum'' are local quantities in the sense
that their value at a point on $\scri$ is constructed from the values
of the field variables at that point. Therefore, these quantities are
not sensitive to the topology of the ``cuts''. In contrast, the Bondi
quantities are global concepts and there is currently no way to
determine their value from only local information. As a consequence
they are very sensitive to the topology of the cuts and, in fact, they
are so sensitive that in our case study with ``cuts'' which have a
toroidal topology there does not exist an energy-momentum
\textit{four}-vector but only one scalar quantity which we still call
the Bondi mass. The reason behind this unexpected phenomenon will be
discussed below.

The main problem one is faced with when trying to obtain expressions
for the asymptotic quantities is the fact that $\scri$ does not look
the way it does in the analytical treatments. In particular, there one
usually assumes that a conformal gauge has been chosen so that $\scri$
is divergence free, i.e., that the area of a cut does not change when
the cut is moved along the null generators of $\scri$. Since $\scri$
itself is shear free this implies that the shape of a cut does not
change either along $\scri$ and this fact can be exploited to choose
the metric on that family of cuts to be one with constant
curvature. Usually this is the unit sphere metric. In our case, where
the cuts have toroidal topology one would choose a flat metric on the
cuts.

In a numerical treatment where the conformal factor $\Omega$ is one of
the evolving variables one has almost no control of its behaviour (at
least at present). Thus, we do not have the freedom to specify that
$\scri$ should have these nice properties and have to live with the
way it emerges from the numerical computation. The only way to
possibly influence the behaviour of the conformal factor is by way of
tuning the gauge source function $\Lambda$ for the conformal
gauge. However, this is a rather indirect way and at the moment it is
completely unclear whether (and how) one should specify $\Lambda$ so
that $\scri$ does have the desired properties.

Another point is that the radiative quantities are referred to a
specific tetrad (or spin frame) on $\scri$ which is adapted to the
geometry there. Again, in the numerical treatment the tetrad is fixed
by other means which implies that we need to transform from the given
tetrad to the geometric tetrad in order to obtain the correct values
of the asymptotic quantities. Again, it is not yet known how to impose
gauge conditions so that the computed tetrad always coincides with the
geometric tetrad on $\scri$. The transformation from the numerical
frame to the adapted frame is straightforward. Recall that the
condition imposed on the adapted spin frame $(O^A,I^A)$ is
\cite{PenroseRindlerII}:
\begin{equation}
  \label{eq:FrameOnScri}
  \nabla_{AA'} \Omega = -A I_A I_{A'} + O(\Omega).
\end{equation}
This condition fixes the direction of the null vector $I^A I^{A'}$ but
says nothing about the space-like vector $O^AI^{A'}$ and its complex
conjugate. Given a cut of $\scri$ these are required to be tangent
vectors to the cut. Then the transformed spin frame is fixed up to the
scalings $O^A\mapsto c\cdot O^A, I^A \mapsto c^{-1}\cdot I^A$.
The transformation to the new spin frame is 
\begin{align}
  O^A &= a o^A + b \iota^A,\\
  I^A &= \frac{1}{a\bar a + b \bar b}(-\bar b o^A  + \bar a \iota^A),
\end{align}
with $a=-c\frac{2\Sigma_{22}}{\Sigma}$ and
$b=c\left(\frac{2\Sigma_{21}}{\Sigma}-1\right)$ and $c$ an arbitrary complex
function on the cut.  With this choice of $a$ and $b$ we have
achieved that on $\scri$ the null vector $I^A I^{A'}$ is aligned with
the null generators of $\scri$. The factor $A$ in
\eqref{eq:FrameOnScri} is found to be 
\begin{equation}
  \label{eq:Afactor}
  A = 4 {c\bar c} \left(\Sigma_{21} - \2\Sigma\right),
\end{equation}
and we furthermore fix $c=\2$ for the remainder of this section.

The asymptotic quantities with respect to the adapted frame can now be
expressed on $\scri$ in terms of the field variables. These expressions
are rather lengthy in the general case but quite manageable in the
symmetry reduced case that we are looking at here. Following is
a list of the variables which are of interest to us and the
expressions to compute them in terms of the field variables in the
reduced case:
\begin{align}
  \sigma'&=0,\quad \kappa'=0,\\
  \rho'&=-\frac{2S}{\Sigma},\quad \tau'=-\frac{S s_{20}}{\Sigma}\\
  \sigma &= -\frac18\left( K_{40} s_{22}^4 + K_{44} + 6 K_{42} s_{22}^2 \right)
 \\
  \psi_2 &= \frac14 \left( E_0 s_{22}^2 - 2 E_2 + E_4 s_{20}^2 \right),\\
  \psi_4 &= E_4 s_{20}^4 + 4\,B_3 s_{20}^3 + 6\,E_2 s_{20}^2
  + 4\,B_1 s_{20} + E_0,\\
  \cN &=\frac14 \left(\phi_{44} s_{20}^4 + 6 \phi_{42} s_{20}^2 +
    \phi_{40}\right)
\end{align}
where we have introduced $s_{22}=\frac{2\Sigma_{22}}{\Sigma}$ and
$s_{20} = \frac{2\Sigma_{20}}{\Sigma}$. The function $\cN$ is the so
called ``news'' function.

Having these expressions at hand it is in principle straightforward to
obtain the asymptotic quantities from the numerical data. The only
obstacle is that the level sets of $\Omega$ do not necessarily agree
with grid lines so that one has to trace out the zero set of $\Omega$
within the grid and then interpolate for the values of the field
variables there. This task is greatly simplified by using the $\scri$
fixing shift gauge discussed in section \ref{sec:gauge} when it is
possible to align $\scri$ on a grid-line initially.

\begin{figure}[htbp]
  \begin{center}
    \leavevmode
    \includegraphics[width=170pt]{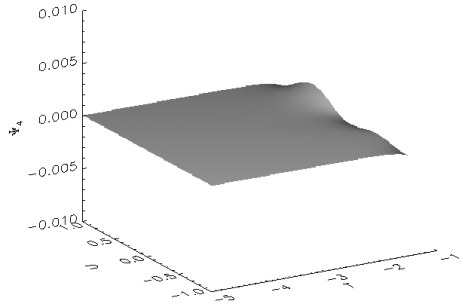}    
    \caption{The ``null datum'' for the A3 solution.}
    \label{fig:A3_Psi4}
  \end{center}
\end{figure}

\begin{figure}[htbp]
  \begin{center}
    \leavevmode
    \includegraphics[width=170pt]{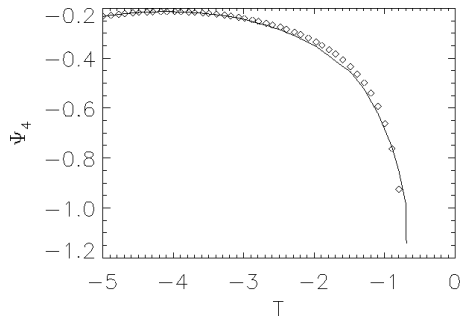}    
    \caption{Time profile of $\psi_4$ for the W1 solution at a
      constant value of $u$.}
    \label{fig:W1_Psi4}
  \end{center}
\end{figure}

In Fig.~\ref{fig:A3_Psi4} we present a surface representation of the null
datum $\psi_4$ for the A3 solution. This is a non-radiating solution
so $\psi_4$ should vanish. Indeed, we find that only when the
singularity is approached the function differs significantly from
zero. This is due to the closeness of the singularity. We should also
point out that this figure has been produced in the warped coordinate
system without the use of the $\scri$ fixing shift gauge. It is only
in the late stages and in the central region where the warping is
maximal when the tracing out of $\scri$ produces too large errors. In
a similar way the W1 solution was treated. Now, $\psi_4$ cannot be
expected to vanish because this is a radiating solution. Since there
is an additional symmetry present in the solution which is aligned
along $\scri$ the function should be constant on $\scri$. We found that the
tracing algorithm works quite satisfactorily in this case also in that
the computed $\psi_4$ is indeed constant as a function of the
$u$-coordinate along $\scri$. Therefore, we show in Fig.~\ref{fig:W1_Psi4}
only a time profile for constant $u$. The line indicates the exact
function while the markers indicate the computed values. The relative
error in this calculation (200 by 200 points) is a few percent in the
region where the influence of the singularity is not too strong.

Let us now discuss some of the issues related to the Bondi
energy-momentum. This is an unexpectedly complicated issue which, in
addition, depends on the global topology of the space-time under
consideration. The standard definition used here is from
\cite{PenroseRindlerII}:
\begin{equation}
  \label{eq:bondimass}
  m_B\left[W\right] = \oint_C \left\{ \psi_2 - \frac{\sigma \cN}{A} 
  \right\} W\; d^2\cS,
\end{equation}
where the integration is over a cut of $\scri$. As it stands the
formula is only valid under rather stringent simplifying
assumptions. It is assumed that the surfaces $\Omega=\mathrm{const}$
are null even away from $\scri$. This implies that $\scri$ is non
diverging and that the spin-coefficient $\tau'$ vanishes. If these
assumptions are not made, then the news $\cN$ acquires additional
compensating terms.

The function $W$ which appears in \eqref{eq:bondimass} is a function
with conformal weight $+1$ on the cut satisfying the conformally
invariant second order elliptic equation
\begin{equation}
  \label{eq:eth2W}
  \eth_c^2 W = 0.
\end{equation}
Here, the $\eth_c$ is the conformally invariant ``eth'' operator
introduced in \cite{PenroseRindlerII}. For a more standard form of
this equation, we refer to \cite{Geroch-1977}. The purpose of solving
the equation \eqref{eq:eth2W} is to select out of the
super-translation subgroup of the asymptotic symmetry group (the BMS
group) the normal subgroup of translations which is used to generate
the energy-momentum. In the special case, where the metric on the cut
has been scaled to be the standard unit sphere metric and where the
cut sits within $\scri$ in such a way that the spin-coefficient $\tau$
vanishes, then the equation \eqref{eq:eth2W} has four linearly
independent solutions which can be taken as the first four spherical
harmonics $Y_{lm},\; l=0,1$. Note, that $\tau=0$
can always be achieved in the neighbourhood of a single cut, because
it only involves parallel transport of $O^A$ along the null
generators. However, given a system of cuts and an adapted spin frame,
this condition cannot, in general, be maintained. Unfortunately, this
is the case for the cuts appearing in the numerical treatment as
intersections of $\scri$ with the constant time hypersurfaces. A more
thorough discussion of the general spherical case is left to another
paper.

Here we want to focus on our immediate interest, namely obtaining a
formula for the Bondi mass on cuts with toroidal topology. In that
case, the BMS group has a completely different structure. This is
reflected in the fact that equation \eqref{eq:eth2W} on a torus has
only a one dimensional solution space as opposed to the four
dimensions in the spherical case. This means that the translation
subgroup is a one-dimensional subgroup of the BMS group. Therefore, on
toroidal cuts, there does not exist a four-vector of energy-momentum,
but only a ``Bondi scalar'', which we call the Bondi mass.

\begin{figure}[htbp]
  \begin{center}
    \leavevmode
    \includegraphics[width=170pt]{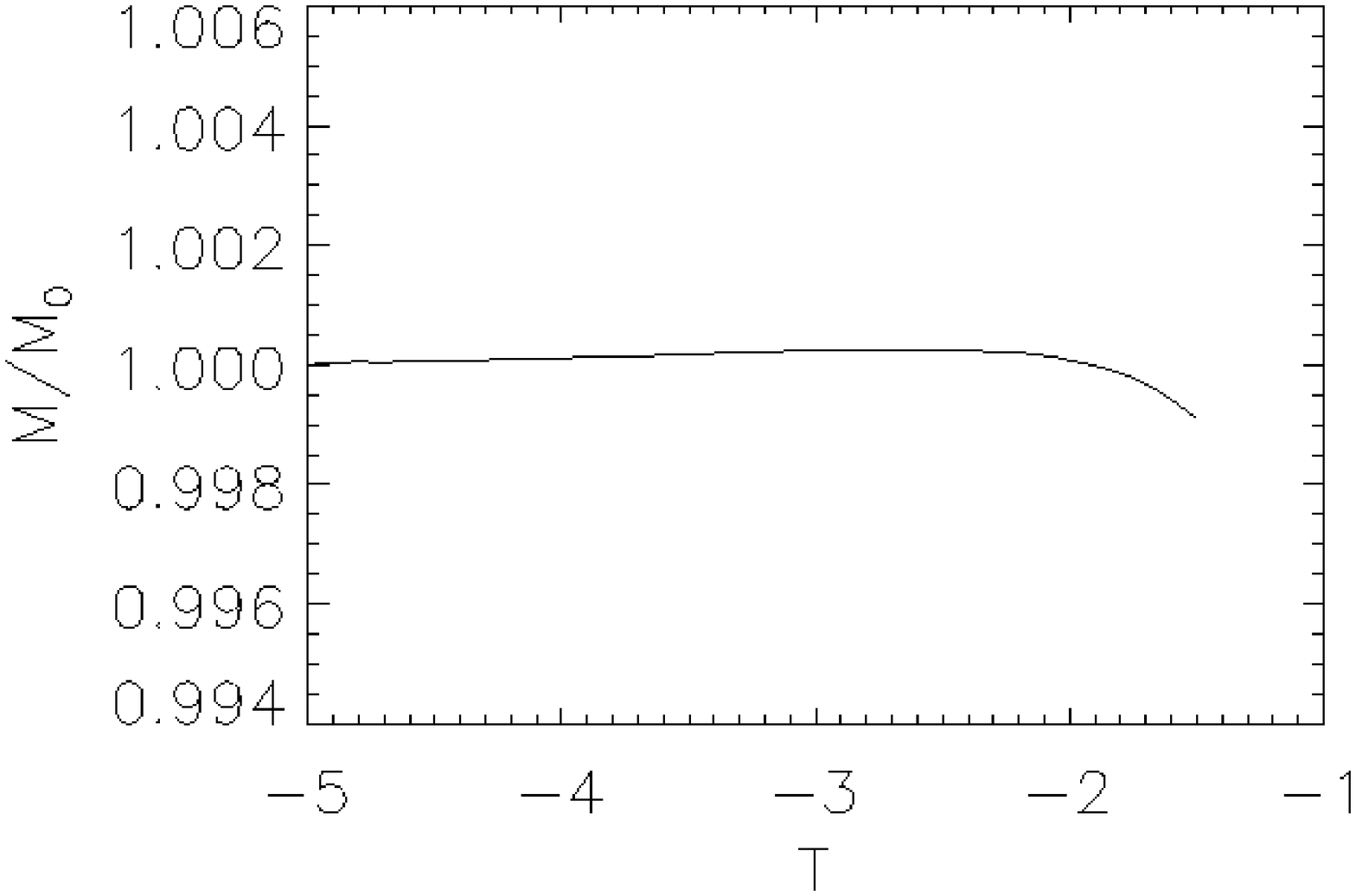}    
    \caption{The Bondi mass for the A3 solution normalised against its
      initial value.}
    \label{fig:A3_Mass}
  \end{center}
\end{figure}

\begin{figure}[htbp]
  \begin{center}
    \leavevmode
    \includegraphics[width=170pt]{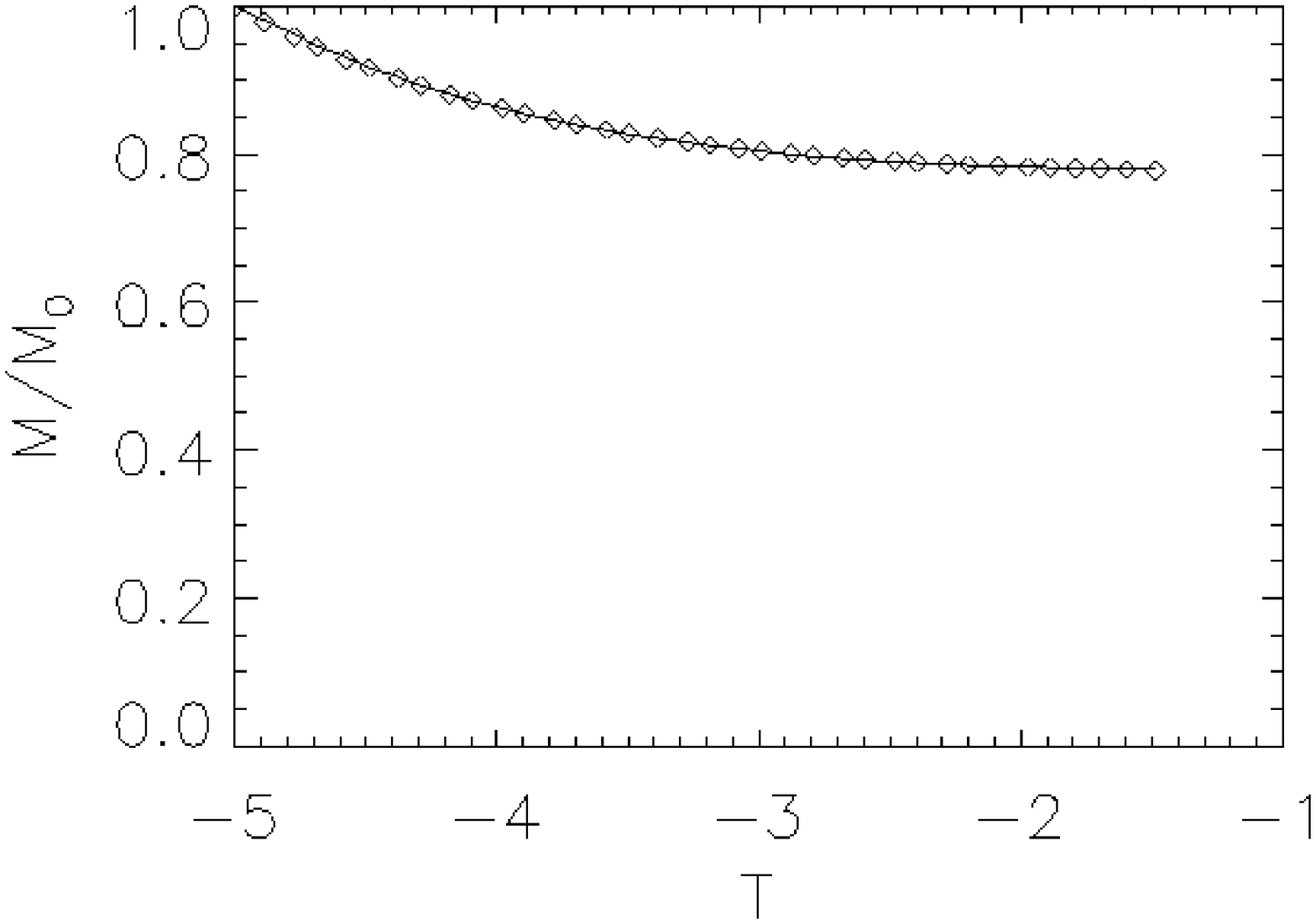}    
    \caption{The Bondi mass for the W1 solution normalised against its
      initial value.}
    \label{fig:W1_Mass}
  \end{center}
\end{figure}

In order to compare the evolution of that scalar with time in our
special case we observe that for the initial data obtained from the
exact solutions A3 and W1, the cuts are spanned by Killing
vectors. This implies that on the cut all field variables are
constant. Hence, we may take $W=\mathrm{const}$ as a solution of
\eqref{eq:eth2W} and since $W$ has to be a conformal density of weight
$+1$ we take $W=\sqrt{\cA}$, with $\cA$ being the area of the
cut. Thus, we end up with the formula
\begin{equation}
  \label{eq:mbtorus}
  m_B = -\sqrt{\cA} \oint_C \left\{ \psi_2 - \frac{\sigma}{A}
    \left(\cN - \bar\sigma \rho' \right)
  \right\}\; d^2\cS.
\end{equation}
Of course, due to the constancy of the integrand on the cut we could
have written this formula without the integral. However, we implement
the formula with the integral because it averages over the numerical
inaccuracies present from the interpolation process. In
Fig.~\ref{fig:A3_Mass} is shown the (normalised) Bondi-mass for the A3
solution, which of course should remain constant. Similarly,
Fig.~\ref{fig:W1_Mass} presents the Bondi mass for the W1 solution. Again,
the solid line is the exact profile while the markers are the values
obtained from the numerical solution.

\section{Conclusion}

In this article we have presented and discussed several issues
concerning the numerical solution of the evolution part of the
hyperboloidal initial value problem for the vacuum conformal field
equations. We have described a special case where the unphysical
assumption was made that there exists a hypersurface orthogonal
Killing vector with closed orbits and no fixed points. This does not
alter the essential issues. The numerical evolution scheme is a simple
two-dimensional implementation of the well-known Lax-Wendroff
method. The outer boundary is evolved using a stable eigen-field
method. We have discussed various lapse choices and the features which
appear when one specifies the gauge source function $F$ as a function
of the field variables. We have found a special choice for the shift
vector which originates in the conformal properties of the
system. This shift allows us to freeze null infinity on the grid while
still leaving the usual freedom for specifying a shift vector in the
interior. Finally, we have described how to obtain the local radiative
information by simply ``reading it off'' $\scri$ and transforming to
the appropriate asymptotic spin frame. The global quantities like
Bondi four-momentum are more difficult to determine and they are very
different in our present case from the physical case where $\scri$ has
spherical sections. We have tested the code and the radiation
extraction algorithm using exact solutions. We obtained good agreement
between the analytical and the numerical solution. Unfortunately, the
used exact solutions have an additional Killing vector which makes
them rather special even though we try to compensate for this by
``warping'' the coordinate system.  Naturally, the next step has to be
to obtain more general initial data.

\begin{appendix}
\section{The exact solutions}
\label{sec:exsol}
We have used several exact solutions for numerical tests. Apart from
the trivial ones which are simply Minkowski space in disguise, i.e.,
rescaled with an arbitrary conformal factor there is the class of
vacuum space-times with toroidal null infinities which have been
constructed by Schmidt \cite{Schmidt-1996} for exactly that
purpose. They are characterised by a solution $w$ of a two-dimensional
wave equation and are defined as follows
\begin{align}
  \label{eq:W}
      \Omega &= \frac18\left(t^2 - z^2\right)\\
      g &= \frac{e^{2n(t,z)}}{\sqrt{t^2 + z^2}} \left(d\,t^2 - d\,z^2\right)
      - \left(t^2 + z^2\right) \left(e^{2w(t,z)} d\,x^2 + e^{-2w(t,z)} 
        d\,y^2\right).
\end{align}
Given a solution of the two-dimensional wave equation \cite{Huebner-1997}
\begin{equation}
  \label{eq:wave}
  \left(t^4-z^4\right) \left( w_{tt}- w_{zz} \right)  
  -2t\left(3z^2+t^2\right)\,w_t
  -2z\left(3t^2+z^2\right)\,w_z = 0,
\end{equation}
one can obtain the function $n(t,z)$ by quadratures. The coordinates
$x$ and $y$ are Killing coordinates, each taking values in $\RR$. We
identify the points $(x,y)$ and $(x+1,y+r)$ to obtain the toroidal
topology. In our applications, we always have $r=1$.  The simplest
solutions of this type are the ones obtained by choosing $w(t,z)=0$
(with $n(t,z)=0$) and $w(t,z) = A(t^2 - v^2)$ for some constant $a$
(with $n(t,z)=-\frac12 A^2(t^2 + z^2)^2$). Note, that $A=0$ in the
latter solution gives the former. The physical metric which
corresponds to the first of these appears in the classification by
Ehlers and Kundt \cite{EhlersKundt-1962} under the name A3 as the
analogue of the Schwarzschild metric in plane symmetry. Here are the
explicit expressions for the variables we use in the code.

\begin{align*}
N&=\frac1{\sqrt{2}}\,{\frac {1}{\sqrt[4]{U}}}{e^{-\frac12\,{A}^{2}U^{2}}},&
K&=\frac{t}{\sqrt{2}}\,\frac { \left (4\,{A}^{2}U^{2} -
      3\right)}{U^{3/4}}{e^{\frac12\,{A}^{2}U^{2}}},\\
{C_{31}}&= \frac1{\sqrt{2}}\,{\frac{1}{\sqrt{U}}}{e^{-A\left(t^2 -
      v^2\right)}},& 
{K_{20}}&=-i\frac{v}{\sqrt{2}}\,{\frac {4\,U^{2}{A}^{2} +
      1}{U^{3/4}}}\,{e^{\frac12\,{A}^{2} U^{2}}},\\ 
{C_{10}}&=\frac1{\sqrt{2}}\,{\frac{1}{\sqrt U}}e^{A\left(t^2 - v^2\right)},&
{K_{40}}&=\frac{t}{2\sqrt{2}}\,{\frac {4\,{A}^{2}U^{2} +
      4\,AU+3}{U^{3/4}}}\,{e^{\frac12\,{A}^{2}U^{2}}},\\
{C_{20}}&=\frac{i}{\sqrt{2}}\,\sqrt[4]U{e^{\frac12\,{A}^{2}U^{2}}},&
{K_{42}}&= \frac{t}{6\sqrt2}\,{\frac {4\,{A}^{2}U^{2}-12\,AU +3}{U^{3/4}}}\,{e^{\frac12\,{A}^{2}U^{2}}},\\
{\gamma_{20}}&=i\frac{v}{\sqrt{2}}\,{\frac{1}{U^{3/4}}}\,{e^{\frac12\,{A}^{2}U^{2}}},&
{\phi_{42}}&=\frac16\,{\frac{4\,{A}^{2}U^{2} - 12\,AU+1}{\sqrt U}}\,{e^{{A}^{2}U^{2}}},\\
{\gamma_{41}}&=-i\frac{v}{\sqrt{2}}\,\sqrt[4]U A{e^{\frac12\,{A}^{2}U^{2}}},&
{\phi_{40}}&=\frac12{\frac{4\,{A}^{2}U^{2}+4\,AU+1}{\sqrt U}}\,{e^{{A}^{2}U^{2}}},\\
F&=4\,{\frac {t}{\sqrt U}}\,{e^{{A}^{2}U^{2}}}.&
{E0}&=-2\,{\frac{8\,U^{3}{A}^{3}+12\,{A}^{2}U^{2} +
      6\,AU-3}{U^{3/2}}}\,{e^{{A}^{2}U^{2}}},\\
S&=\frac14\,{e^{{A}^{2}U^{2}}}\sqrt U,&
{E2}&=2\,{\frac {8\,U^{3}{A}^{3} -
      4\,{A}^{2}U^{2}+6\,AU+1}{U^{3/2}}}\,{e^{{A}^{2}U^{2}}},\\
\Sigma&=\frac1{2\sqrt2}\,t\sqrt[4]U{e^{\frac12\,{A}^{2}U^{2}}},&
\phi&=-2\,{\frac{4\,{A}^{2}U^{2} + 1}{\sqrt U}}\,{e^{{A}^{2}U^{2}}},\\
&&{\Sigma_{20}}&=-i \frac{v}{2\sqrt2}\,\sqrt[4]U{e^{\frac12\,{A}^{2}U^{2}}},
\end{align*}
with $U=t^2+z^2$. All other functions either vanish or they are
complex conjugates of functions given above.
\end{appendix}

\section*{Acknowledgements}

It is a pleasure for me to thank the members of the Mathematical
Relativity group at the Max-Planck-Institut f\"ur Gravitationsphysik
in Potsdam where part of this work has been done.

\newpage
\section*{Figures}


\begin{thebibliography}{10}

\bibitem{jf-1997-2}
J.~Frauendiener.
\newblock {\sl Numerical treatment of the hyperboloidal initial value problem
  for the vacuum Einstein equations. I. The conformal field equations}.
\newblock preprint.

\bibitem{Schmidt-1996}
B.~G. Schmidt.
\newblock {\sl Vacuum space-times with toroidal null infinities}.
\newblock {\em Class. Quant. Grav.}, {\bf 13}, p.~2811--2816, 1996.

\bibitem{PenroseRindlerII}
R.~Penrose and W.~Rindler.
\newblock {\em Spinors and Spacetime}, volume~2.
\newblock Cambridge University Press, 1986.

\bibitem{McLenaghan-1987}
R.~McLenaghan.
\newblock {\sl NP: A Maple package for performing calculations in the
  {N}ewman-{P}enrose formalism}.
\newblock {\em Gen. Rel. Grav.}, {\bf 19}, p.~623--635, 1987.

\bibitem{CourantFriedrichsLewy-1928}
R.~Courant, K.~O. Friedrichs, and H.~Lewy.
\newblock {\sl {\"U}ber die partiellen {D}ifferenzengleichungen der
  mathematischen Physik}.
\newblock {\em Math. Ann}, {\bf 100}, p.~32--74, 1928.

\bibitem{Kreiss-1968}
H.~O. Kreiss.
\newblock {\sl Stability theory for difference approximations of mixed initial
  boundary value problems. I}.
\newblock {\em Math. Comp.}, {\bf 22}, p.~703--714, 1968.

\bibitem{GustafssonKreissSundstroem-1972}
B.~Gustafsson, H.~O. Kreiss, and A.~Sundstr{\"o}m.
\newblock {\sl Stability theory for difference approximations of mixed initial
  boundary value problems. II}.
\newblock {\em Math. Comp.}, {\bf 26}, p.~649--686, 1972.

\bibitem{Trefethen-1982}
L.~N. Trefethen.
\newblock {\sl Group velocity in finite difference schemes}.
\newblock {\em SIAM Review}, {\bf 24}, p.~113--136, 1982.

\bibitem{FriedrichNagy-1997}
H.~Friedrich and G.~Nagy.
\newblock {\sl The initial boundary value problem for Einstein's vacuum field
  equations}.
\newblock preprint, 1998.

\bibitem{LeVeque-1997}
R.~J. LeVeque.
\newblock {\sl Wave propagation algorithms for multidimensional hyperbolic
  systems}.
\newblock {\em J. Comp. Phys.}, {\bf 131}, p.~327--353, 1997.

\bibitem{Geroch-1977}
R.~Geroch.
\newblock {\sl Asymptotic structure of space-time}.
\newblock In F.~P. Esposito and L.~Witten, editors, {\em Asymptotic structure
  of space-time}. Plenum, New York, 1977.

\bibitem{Huebner-1997}
P.~H{\"u}bner.
\newblock {\sl More about Vacuum Spacetimes with Toroidal Null Infinities}.
\newblock unpublished, 1997.

\bibitem{EhlersKundt-1962}
J.~Ehlers and W.~Kundt.
\newblock {\sl Exact solutions of the gravitational field equations}.
\newblock In L.~Witten, editor, {\em Gravitation: an introduction to current
  research}. J. Wiley, New York, 1962.

\end{thebibliography}
\end{document}